# Nonlinear higher-order hydrodynamics: Fluids under driven flow and shear pressure

Clóves G. Rodrigues,[1,a)]  José G. Ramos,[2] Carlos A. B. Silva,[3] and Roberto Luzzi[2]

[1]School of Exact Sciences and Computing, Pontifical Catholic University of Goiás, CP 86, 74605-010 Goiânia, Goiás, Brazil
[2]Condensed Matter Physics Department, Institute of Physics "Gleb Wataghin," State University of Campinas-Unicamp, 13083-859 Campinas, SP, Brazil
[3]Departamento de Física, Instituto Tecnológico de Aeronáutica, 12228-901 São José dos Campos, SP, Brazil

[a)]Author to whom correspondence should be addressed: cloves@pucgoias.edu.br

**ABSTRACT**

In the context of a nonequilibrium statistical thermodynamics—based on a nonequilibrium statistical ensemble formalism—a generalized hydrodynamics of fluids under driven flow and shear stress is derived. At the thermodynamic level, the nonequilibrium equations of state are derived, which are coupled to the evolution of the basic variables that describe the hydrodynamic motion in such a system. Generalized diffusion-advection and Maxwell–Cattaneo advection equations are obtained in appropriate limiting situations. This nonlinear higher-order hydrodynamics is applied, in an illustration, to the case of a dilute solution of Brownian particles in nonequilibrium conditions and flowing in a solvent acting as a thermal bath. This is done in the framework of such generalized hydrodynamics but truncated up to a second order.

## I. INTRODUCTION

The thermodynamics of fluids under shear flow is an active and very challenging topic in modern nonequilibrium thermodynamics and statistical mechanics. The field has an appealing interest for involving not only the purely physical aspects, but also in what concerns technological–industrial applications. Theoretically, it encompasses what can be termed as thermo-hydrodynamics because it requires an open discussion involving simultaneously thermodynamics and hydrodynamics. It can also be noticed that the formulation of a nonequilibrium thermodynamics and hydrodynamics, with an interplay between the macroscopic and microscopic levels of description, via a nonequilibrium statistical mechanics, currently represents a fundamental frontier in the physical sciences.

Several phenomena induced by flow in fluid systems have been reviewed and analyzed in the book.[1] It is our aim to reconsider, and extend, most of the questions involved in the framework of a thermo-statistics for dissipative systems which can be dubbed as informational statistical thermodynamics[2,3] (IST). Indeed, a close interaction between macroscopic and microscopic approaches is necessary and convenient. On the one hand, it gives explicit expressions, derived from the microscopic description. On the other hand, the macroscopic approach outlines some common features that should be shared by very different physical systems.

In the present paper, we explore the thermo-hydrodynamics of a classical fluid under flow and shear stress and in contact with a thermal bath, resorting, as noticed, to a treatment in the context of IST. Thus, it has a microscopic (mechanical–statistical) background and nonlocality and nonlinearity effects can be incorporated.

It has been noticed[4] that a nonlinear flow behavior is fairly common in complex fluids consisting of substances of biological and of technical importance, such as, polymeric liquids, surfactant solutions, and colloidal dispersions. Also,[5] nonlinearity includes the possibility of emergence of singularities in finite time.

Informational-statistical thermodynamics was initiated by Hobson[6,7] a few years after the publication of Jaynes seminal papers[8,9] on the foundation of statistical mechanics on information theory. IST was later on systematized and extended as described in Refs. 2 and 3. This statistical thermodynamics is based on a nonequilibrium statistical ensemble formalism (NESEF) of a quite general scope,[10–13] and, within it, on a particular version consisting of a nonequilibrium grand-canonical ensemble, in a mesoscopic description.

In the present work, we use nonequilibrium statistical ensemble formalism (NESEF), which is based on the nonequilibrium statistical operator developed by Zubarev. The method permits one to generalize the Gibbs ensemble method to the nonequilibrium case and to construct a nonequilibrium statistical operator, which enables one to obtain the generalized hydrodynamics of fluids under driven flow. We emphasize that there are other approaches that have been proposed for treating nonlinear hydrodynamics.[14–17] Eu, for instance, uses an ensemble methods on the basis of the Boltzmann equation, the generic Boltzmann equations for classical and quantum dilute gases, and a generalized Boltzmann equation for dense simple fluids.[18,19]

The paper is organized as follows: In Sec. II is described the characterization of the macrostate of the system appropriate for a construction of its thermodynamics. A reduced set of basic variables deemed appropriate for such descriptions in a lower-order hydrodynamics is introduced, and the nonequilibrium statistical distribution is built. In the process is introduced the set of nonequilibrium thermodynamic variables that the formalism defines. It is established the relation of them with the basic variables, in what constitutes a set of nonequilibrium equations of state. Such relations are fundamental for the closing of the hydrodynamics equations of evolution (which are presented in Sec. IV). In Sec. III, the information-theoretic entropy, its rate of change, and the surrogates of free energy and chemical potential in nonequilibrium conditions (usually referred-to as quasi-free energy and quasi-chemical potential) are derived and discussed. The evolution equations, as noticed, are presented in Sec. IV: they are those for the density, the density-flux (related to the linear momentum density and to the barycentric velocity), and the second-density flux (related to the pressure tensor field). In particular limiting cases, namely, generalized (in this level of description) diffusion-advection and Maxwell–Cattaneo advection equations are obtained. Finally, the results described in those sections are applied to the study of the steady state of a planar Couette flow, followed by applications to a planar extensional one and the case of purely shear flow. In Sec. V, we present a brief summary of the subject and additional comments.

## II. NONLINEAR HIGHER-ORDER HYDRODYNAMICS (BRIEF DESCRIPTION)

Let us consider a system of particles of mass $m$ and embedded in a fluid (of molecules of mass $M$), acting as a thermal bath, which is in equilibrium with an external reservoir. An external force acts on the molecules of the system, causing it to flow and which is then driven out of equilibrium. A relate gradient is produced given rise to a shear stress, that is, as noticed in the Introduction, the situation to be studied in detail. There it was also noticed that a kinetic (at the microscopic level) foundation of the hydrodynamics (at the macroscopic level) can be obtained resorting to the nonequilibrium statistical ensemble formalism (NESEF).

For that purpose, the description of the nonequilibrium state of the system is done, in the context of NESEF[10] in terms of the one- and two-particle dynamical operators (what is equivalent to incorporate all observables of the system),[11–13] which at the classical mechanical description are given by

$$\hat{n}_1(\mathbf{r}, \mathbf{p}, \Gamma) = \sum_{j=1}^{N} \delta(\mathbf{r} - \mathbf{r}_j)\delta(\mathbf{p} - \mathbf{p}_j), \quad (1)$$

$$\hat{n}_2(\mathbf{r}, \mathbf{p}, \mathbf{r}', \mathbf{p}'|\Gamma) = \sum_{j \neq k=1}^{N} \delta(\mathbf{r} - \mathbf{r}_j)\delta(\mathbf{p} - \mathbf{p}_j)\delta(\mathbf{r}' - \mathbf{r}_k) \times \delta(\mathbf{p}' - \mathbf{p}_k), \quad (2)$$

where $\Gamma$ stands for a point in phase space, $\mathbf{r}$, and $\mathbf{p}$ stands for the coordinates and linear moments of each particle of mass $m$.

At this point, we simplified matters considering that we are in the presence of a dilute solute (the molecules of mass $m$) in a solvent (the fluid of molecules of mass $M$), and then the interaction between the molecules in the solute can be neglected. Thus, correlations between them can be ignored, that is, we can omit $\hat{n}_2$ in the description to work only with the single-particle dynamical operator $\hat{n}_1$. However, we emphasize that this term, $\hat{n}_2$, may be important. The two-particle contribution may be of particular interest when the interactions between the particles are relevant and may led to a phase transition, for example, interacting Brownian particles could have a transition to an aggregated phase, for instance, a gel state. This is a field of intense research, where both fundamental and applied questions are equally important. Applications include different areas, such as ceramics processing, cosmetics and consumer products, biomaterials, drug delivery systems, food science, etc.[20–24]

Hence, in the framework of NESEF, we need to construct the nonequilibrium statistical operator $\varrho_\varepsilon(t)$ in terms of $\hat{n}_1$. According to the theory,[10–12] it is in this case given by

$$\rho_\varepsilon(t) = \exp\left\{-\hat{S}(t,0) + \int_{-\infty}^{t} dt' e^{-\varepsilon(t'-t)} \frac{d}{dt'}\hat{S}(t',t'-t)\right\}, \quad (3)$$

with $\varepsilon$ being a positive infinitesimal that is taken in the limit of go in to $+0$ after the trace operation has been performed (it accounts for introducing historicity, and irreversibility in the theory,[10,25] and where

$$\hat{S}(t,0) = -\ln \bar{\varrho}(t,0), \quad (4)$$

$$\hat{S}(t',t'-t) = e^{i(t'-t)\varepsilon}\hat{S}(t,0), \quad (5)$$

and

$$\bar{\varrho}(t,0) = \exp\left\{-\phi(t) - \int d^3r d^3p \, \varphi_1(\mathbf{r},\mathbf{p};t)\hat{n}_1(\mathbf{r},\mathbf{p})\right\}, \quad (6)$$

is the so-called auxiliary (or "instantaneously frozen quasi-equilibrium") statistical operator; for simplicity, we have omitted to explicitly write the dependence on the phase point $\Gamma$. In Eq. (4), $\hat{S}(t,0)$ is the so-called NESEF-entropy operator, analyzed in Ref. 26; in Eq. (6), $\phi(t)$, playing the role of the logarithm of nonequilibrium partition function, say,

$$\phi(t) = \ln \bar{Z}(t),$$

$$\phi(t) = \ln \int dt' \exp\left\{-\int d^3r d^3p \, \varphi_1(\mathbf{r},\mathbf{p};t)\hat{n}_1(\mathbf{r},\mathbf{p})\right\}, \quad (7)$$

ensures the normalization of the statistical probability distribution, and $\varphi_1(\mathbf{r},\mathbf{p};t)$ is the nonequilibrium thermodynamic variable said conjugated to the basic variable $\hat{n}_1(\mathbf{r},\mathbf{p},\Gamma)$.

The average value of the basic dynamic quantity, $\hat{n}_1$, namely,

$$f(\mathbf{r},\mathbf{p};t) = \text{Tr}\{\hat{n}_1(\mathbf{r},\mathbf{p})\varrho_\varepsilon(t)\}, \quad (8)$$

can be considered as a generalized Boltzmann single-particle distribution function, which in the framework of the NESEF-based kinetic theory[10,27,28] satisfies.[29,30]

But, for our purposes here of building a generalized nonlinear thermo-hydrodynamics (that is, the thermal physics of continuous media), it is highly convenient to resort to an equivalent nonequilibrium mesoscopic-level in what can be referred-to as a grand-canonical ensemble. This is described elsewhere.[10,31] Here, we simply noticed that, on the one hand, it follows once we redefine the nonequilibrium thermodynamics variable $\varphi$, in the form

$$\varphi_1(\mathbf{r}, \mathbf{p}; t) = A_n(\mathbf{r}, t) + F_h(\mathbf{r}, t)\frac{p^2}{2m} + \mathbf{F}_n(\mathbf{r}, t)$$
$$\cdot \frac{\mathbf{p}}{m} + \mathbf{F}_h(\mathbf{r}, t) \cdot \frac{\mathbf{p}}{2m}\frac{\mathbf{p}}{m}$$
$$+ \sum_{\ell \geqslant 2}\left[F_n^{[\ell]}(\mathbf{r}, t) \otimes u^{[\ell]}(\mathbf{p}) + F_h^{[\ell]}(\mathbf{r}, t) \otimes u^{[\ell]}(\mathbf{p})\right], \quad (9)$$

where upper index $[\ell]$ indicates a rank-$\ell$ tensor, $\otimes$ stands for fully contracted product of tensors, and

$$u^{[\ell]}(\mathbf{p}) = \left[\frac{\mathbf{p}}{m} \cdots (\ell - \text{times}) \cdots \frac{\mathbf{p}}{m}\right] \quad (10)$$

is a rank-$\ell$ tensor resulting from the tensorial inner products, $[\cdots]$, of $\ell$-times the velocity $\mathbf{p}/m$. Moreover, Eq. (9) introduces the set of nonequilibrium thermodynamics variables,

$$\left\{F_h(\mathbf{r}, t), A_n(\mathbf{r}, t), \mathbf{F}_h(\mathbf{r}, t), \mathbf{F}_n(\mathbf{r}, t), \{F_h^{[\ell]}(\mathbf{r}, t)\}, \{F_n^{[\ell]}(\mathbf{r}, t)\}\right\}, \quad (11)$$

which are said to be conjugated to the resulting basic set of dynamical variables consisting of

$$\left\{\hat{h}(\mathbf{r}), \hat{n}(\mathbf{r}), \hat{\mathbf{I}}_h(\mathbf{r}), \hat{\mathbf{I}}_n(\mathbf{r}), \{\hat{I}_h^{[\ell]}(\mathbf{r})\}, \{\hat{I}_n^{[\ell]}(\mathbf{r})\}\right\} \quad (12)$$

given in Eqs. (13)–(18) below. On the other hand, it can be noticed that this can be considered as introducing a far-reaching generalization of Grad's moments approach used for dealing with the standard Boltzmann equation, and then here, in a sense, consists in generalization of it for dealing with the NESEF-based Boltzmann equation in nonequilibrium conditions.

In this generalized nonequilibrium grand-canonical ensemble, the set of dynamical variables in (12), consisting, respectively, of the density of energy, density of particles, the flux (current) of energy, the flux (current) of particles, the set of higher-order ($\ell \geqslant 2$) fluxes of energy, and the set of higher-order ($\ell \geqslant 2$) flux of particles, are

$$\hat{h}(\mathbf{r}) = \int d^3p \frac{p^2}{2m} \hat{n}_1(\mathbf{r}, \mathbf{p}), \quad (13)$$

$$\hat{n}(\mathbf{r}) = \int d^3p\, \hat{n}_1(\mathbf{r}, \mathbf{p}), \quad (14)$$

$$\hat{\mathbf{I}}_h(\mathbf{r}) = \int d^3p \frac{p^2}{2m} \mathbf{u}(\mathbf{r} \cdot \mathbf{p}) \hat{n}_1(\mathbf{r}, \mathbf{p}), \quad (15)$$

$$\hat{\mathbf{I}}_n(\mathbf{r}) = \int d^3p \frac{\mathbf{p}}{m} \hat{n}_1(\mathbf{r}, \mathbf{p}), \quad (16)$$

$$\hat{I}_h^{[\ell]}(\mathbf{r}) = \int d^3p \frac{p^2}{2m} u^{[\ell]}(\mathbf{p}) \hat{n}_1(\mathbf{r}, \mathbf{p}), \quad (17)$$

$$\hat{I}_n^{[\ell]}(\mathbf{r}) = \int d^3p\, u^{[\ell]}(\mathbf{p}) \hat{n}_1(\mathbf{r}, \mathbf{p}), \quad (18)$$

with $u^{[\ell]}(\mathbf{p})$ defined in Eq. (10), and we recall, $\ell = 2, 3, \ldots$.

Considering the purpose of the work, study of fluids under flow and shear stress, we can introduce a truncation of description consisting of retaining the reduced set of basic dynamical variables composed of

$$\left\{\hat{h}(\mathbf{r}), \hat{n}(\mathbf{r}), \hat{\mathbf{I}}_n(\mathbf{r}), \hat{I}_n^{[2]}(\mathbf{r})\right\}, \quad (19)$$

that is, the density of energy, the density of particles, and the first two fluxes of the latter. The criterion and the meaning of this kind of truncation are discussed in Ref. 32, and we notice the use of the whole set of variables, namely, the densities and their fluxes of all order lead to the construction of a higher-order thermo-hydrodynamics based on NESEF (described in Ref. 33) where a generalized Fick's law is derived including a generalized thermodynamic force and the presence of a continuous fraction that accounts for the series of the so-called Burnett and super-Burnett contributions.[34–36] Moreover, Bobylev's instability is characterized and discussed. Therefore, the description in terms of the set of variables in Refs. 37–39 implies in dealing with movements with predominance of long wavelengths.

In terms of the set of dynamical variables in (19), the auxiliary nonequilibrium statistical operator of Eq. (4) becomes

$$\bar{\varrho}(t, 0) = \exp\left\{-\phi(t') - \int d^3r\, \Psi(\mathbf{r}, t)\right\}, \quad (20)$$

where

$$\Psi(\mathbf{r}, t) = F_h(\mathbf{r}, t)h(\mathbf{r}) + A_n(\mathbf{r}, t)n(\mathbf{r}) + \mathbf{F}_n(\mathbf{r}, t) \cdot \mathbf{I}_n(\mathbf{r})$$
$$+ F_n^{[2]}(\mathbf{r}, t) \otimes I_n^{[2]}(\mathbf{r}). \quad (21)$$

We recall that all the dynamical variables depend on the phase space point $\Gamma$ and that $\phi(t)$ ensures the normalization of the auxiliary probability distribution, i.e.,

$$\phi(t) = \ln \int d\Gamma \exp\left\{-\int d^3r\, \Psi(\mathbf{r}, t)\right\}, \quad (22)$$

where, for short, we have written $d\Gamma$ for

$$d\Gamma = \sum_N \prod_j \frac{d^3r_j d^3p_j}{(2\pi\hbar)^{3N} N!}, \quad (23)$$

with the quantity $(2\pi\hbar)^{3N} N!$ being the element of volume in phase space for a system of $N$ indistinguishable particles. Furthermore, we recall that, first, the probability distribution of the system of interest is the on of Eq. (4) using for $\bar{\varrho}$ the expression of Eq. (7), second, that the probability distribution of system and thermal bath is

$$\mathcal{R}_\varepsilon(t') = \varrho_\varepsilon(t') \times \rho_B, \quad (24)$$

where

$$\rho_B = \frac{1}{Z(T_0, N_B, V_B)e^{\hat{H}_B/k_B T_0}} \quad (25)$$

is the canonical distribution of the bath at temperature $T_0$. Finally, the set of nonequilibrium thermodynamics variables conjugated to the set of basic dynamical variables of Eq. (19) consists of [cf. Eqs. (22), (9), and (11)]

$$\{F_h(\mathbf{r}, t'), A_n(\mathbf{r}, t'), \mathbf{F}_n(\mathbf{r}, t'), F_n^{[2]}(\mathbf{r}, t')\}, \quad (26)$$

and the set of basic thermo-hydrodynamic variables consists of the average values over the nonequilibrium ensemble of the set of basic dynamical variables in Eq. (19), namely,

$$h(\mathbf{r},t) = \int d\Gamma \hat{h}(\mathbf{r})\varrho_\varepsilon(t) = \int d\Gamma \hat{h}(\mathbf{r})\bar{\varrho}(t,0), \quad (27)$$

$$n(\mathbf{r},t) = \int d\Gamma \hat{n}(\mathbf{r})\varrho_\varepsilon(t) = \int d\Gamma \hat{n}(\mathbf{r})\bar{\varrho}(t,0), \quad (28)$$

$$\mathbf{I}_n(\mathbf{r},t) = \int d\Gamma \hat{\mathbf{I}}(\mathbf{r})\varrho_\varepsilon(t) = \int d\Gamma \hat{\mathbf{I}}(\mathbf{r})\bar{\varrho}(t,0), \quad (29)$$

$$I_n^{[2]}(\mathbf{r},t) = \int d\Gamma \hat{I}_n^{[2]}(\mathbf{r})\varrho_\varepsilon(t) = \int d\Gamma \hat{I}_n^{[2]}(\mathbf{r})\bar{\varrho}(t,0), \quad (30)$$

where we have indicated that for the basic variables, and only for the basic variables, the average value with the probability distribution $\varrho_\varepsilon(t)$ coincides with the one calculated with the auxiliary one $\bar{\varrho}(t,0)$.[10–12]

We stress that the set of nonequilibrium thermodynamic variables of Eq. (26) define a space of nonequilibrium thermodynamic states that describes the macroscopic state of the system as does the set of basic variables of Eqs. (27)–(30). These equations relate both types of nonequilibrium thermodynamics variables and, thus, Eqs. (27)–(30) can be considered as nonequilibrium equations of state (in analogy to the case in equilibrium). Also, to keep a kind of analogy with the grand-canonical statistical distribution in the approximation thermodynamics variables as

$$F(\mathbf{r},t) = \beta^*(\mathbf{r},t) = \frac{1}{k_B T^*(\mathbf{r},t)}, \quad (31)$$

introducing a nonequilibrium local and time-dependent temperature, $T^*(\mathbf{r},t)$, which is referred-to as quasi-temperature (or nonequilibrium temperature),[37–41]

$$A_n(\mathbf{r},t) = -\beta^*(\mathbf{r},t)\mu^*(\mathbf{r},t) = -\frac{\mu^*(\mathbf{r},t)}{k_B T^*(\mathbf{r},t)}, \quad (32)$$

introducing a local and time-dependent chemical potential, $\mu^*(\mathbf{r},t)$, which is referred-to as quasi-chemical potential,[37–42] and

$$\mathbf{F}_n(\mathbf{r},t) = -\beta^*(\mathbf{r},t)m\mathbf{v}_n(\mathbf{r},t) = -\frac{m\mathbf{v}_n(\mathbf{r},t)}{k_B T^*(\mathbf{r},t)}, \quad (33)$$

introducing a drift velocity, $\mathbf{v}_n(\mathbf{r},t)$ associated with the flux. Moreover, we write

$$F_n^{[2]}(\mathbf{r},t) = -[\beta^*(\mathbf{r},t)m]^2 v_n^{[2]}(\mathbf{r},t) = -\frac{m^2 v_n^{[2]}(\mathbf{r},t)}{[k_B T^*(\mathbf{r},t)]^2}, \quad (34)$$

introducing the second-rank tensor, $v_n^{[2]}(\mathbf{r},t)$, having dimensions of a square velocity, for the nonequilibrium thermodynamic variable associated with the second order flux [the latter is related to the pressure tensor as shown below: cf. Eq. (42)]. The question of the definition of a space of thermodynamic states has been considered by Meixner in relevant papers, which apparently did not receive appropriate attention.[43,44]

At this point, we recall that the generalized nonequilibrium grand-canonical ensemble we are using includes two families of basic thermodynamic variables [cf. Eqs. (13)–(18)]: one associated with the thermal motion (the density of energy and its fluxes) and the other associated with material motion (the density of particles and its fluxes), which are composed of linear combinations of the one-particle dynamical operator $\hat{n}_1$. This introduces a partial redundancy in the description in the sense of existing partial relationships between the two families, which is eliminated imposing some restrictions, in the present case consisting in that

$$F_n^{[2]}(\mathbf{r},t) \otimes \mathbf{1}^{[2]} - \text{Tr}\{F_n^{[2]}(\mathbf{r},t)\} = 0, \quad (35)$$

i.e., the tensor is traceless.[10]

In terms of these redefinitions, the set of nonequilibrium thermodynamical variables indicated in ((26) consists of

$$\{\beta^*(\mathbf{r},t), \mu^*(\mathbf{r},t), \mathbf{v}_n(\mathbf{r},t), v_n^{[2]}(\mathbf{r},t)\}. \quad (36)$$

Using Eqs. (31)–(34) in Eq. (20), we obtain that

$$\bar{\varrho}(t,0) = \exp\left\{-\int d^3 r\, \hat{s}(\mathbf{r},t)\right\}, \quad (37)$$

where

$$\hat{s}(\mathbf{r},t) = \beta^*(\mathbf{r},t)\Omega(\mathbf{r},t) + \beta^*(\mathbf{r},t)[\hat{h}(\mathbf{r}) - \mu^*(\mathbf{r},t)\hat{n}(\mathbf{r})$$
$$- m\mathbf{v}_n(\mathbf{r},t) \cdot \mathbf{I}_n(\mathbf{r},t) - m^2\beta^*(\mathbf{r},t)v_n^{[2]}(\mathbf{r},t) \otimes I_n^{[2]}(\mathbf{r},t)], \quad (38)$$

and we have introduced the definition

$$\phi(t) = -\int d^3 r\, \beta^*(\mathbf{r},t)\Omega(\mathbf{r},t), \quad (39)$$

with $\Omega(\mathbf{r},t)$ playing the role of a nonequilibrium grand-canonical free energy density, and we recall that $v_n^{[2]}(\mathbf{r},t)$ is a traceless second-rank tensor [cf. Eqs. (35) and (34)]. In terms of set of variables in Eq. (36), it follows that [cf. Eqs. (22)–(34)—redefinition of nonequilibrium thermodynamical variables—and Appendix A]

$$\psi(t) = \ln \int d\Gamma \exp\left\{-\int d^3 r\beta^*(\mathbf{r},t)\Big[\hat{h}(\mathbf{r},t) - \mu^*(\mathbf{r},t)\hat{n}(\mathbf{r})\right.$$
$$\left. - m\mathbf{v}_n(\mathbf{r},t) \cdot \mathbf{I}(\mathbf{r}) - m^2\beta^*(\mathbf{r},t)v_n^{[2]}(\mathbf{r},t) \otimes I_n^{[2]}(\mathbf{r},t)\Big]\right\}, \quad (40)$$

$$\psi(t) = \left(\frac{2\pi m}{\hbar^2}\right)^{3/2} \int d^3 r \frac{1}{\sqrt{M(\mathbf{r},t)}} \exp\left\{-\beta^*(\mathbf{r},t)\mu^*(\mathbf{r},t)\right.$$
$$\left. + \frac{1}{2}mM^{[2]}(\mathbf{r},t) \otimes [\mathbf{v}(\mathbf{r},t)\mathbf{v}(\mathbf{r},t)]\right\},$$

where $M(\mathbf{r},t)$ is the determinant of the matrix,

$$M^{[2]}(\mathbf{r},t) = \beta^*(\mathbf{r},t)\mathbf{1}^{[2]} + \frac{2}{m}\mathcal{F}_n^{[2]}(\mathbf{r},t) = 0, \quad (41)$$

and $\mathbf{v}(\mathbf{r},t)$ is the barycentric velocity field [cf. Eq. (B15)], $\mathbf{1}^{[2]}$ is the unit rank-2 tensor and we recall that $A_n(\mathbf{r},t) = -\beta^*(\mathbf{r},t)\mu^*(\mathbf{r},t)$.

At this point, we introduce the pressure tensor field $P^{[2]}(\mathbf{r},t)$, which is the flux of the linear momentum in the barycentric coordinates, that is,

$$P^{[2]}(\mathbf{r},t) = mI_n^{[2]}(\mathbf{r},t) - n(\mathbf{r},t)m[\mathbf{v}(\mathbf{r},t)\mathbf{v}(\mathbf{r},t)],$$
$$P^{[2]}(\mathbf{r},t) = n(\mathbf{r},t)\mathcal{M}^{[2]}(\mathbf{r},t), \quad (42)$$
$$P^{[2]}(\mathbf{r},t) \equiv p(\mathbf{r},t)\mathbf{1}^{[2]} + \mathring{P}^{[2]}(\mathbf{r},t),$$

where $\mathcal{M}^{[2]}$ is the inverse of $M^{[2]}$, and we have introduced

$$p(\mathbf{r},t) = m\frac{1}{3}\mathrm{Tr}\{I_n^{[2]}(\mathbf{r},t)\} - n(\mathbf{r},t)\frac{m}{3}v^2(\mathbf{r},t) = \frac{2}{3}u(\mathbf{r},t), \quad (43)$$

which is the kinetic pressure field, after using that

$$h(\mathbf{r},t) = u(\mathbf{r},t) + n(\mathbf{r},t)\frac{1}{2}mv^2(\mathbf{r},t), \quad (44)$$

with

$$u(\mathbf{r},t) = \frac{1}{2}n(\mathbf{r},t)\mathrm{Tr}\{\mathcal{M}^{[2]}(\mathbf{r},t)\} = \frac{3}{2}p(\mathbf{r},t) \quad (45)$$

being the internal energy and the last term in Eq. (44) is the kinetic energy of drift.

Finally,

$$\mathring{P}^{[2]}(\mathbf{r},t) = P^{[2]}(\mathbf{r},t) - \frac{1}{3}\mathrm{Tr}\{P^{[2]}(\mathbf{r},t)\}\mathbf{1}^{[2]},$$
$$\mathring{P}^{[2]}(\mathbf{r},t) = n(\mathbf{r},t)\left[\mathcal{M}^{[2]}(\mathbf{r},t) - \frac{1}{3}\mathrm{Tr}\{\mathcal{M}^{[2]}(\mathbf{r},t)\}\mathbf{1}^{[2]}\right] \quad (46)$$

is the traceless contribution to the pressure tensor field.

Using the nonequilibrium equations of state given in Appendix A [cf. Eqs. (A11)–(A16)], and the equations above, we can obtain the following relations between the basic variables with the pressure tensor and nonequilibrium thermodynamic variables, that is, they are the nonequilibrium equations of state in the description introducing the pressure tensor as alternative basic variables in place of the second flux,

$$h(\mathbf{r},t) = \mathrm{Tr}\{P^{[2]}(\mathbf{r},t)\} + n(\mathbf{r},t)\frac{1}{2}mv^2(\mathbf{r},t), \quad (47)$$

or, after using Eqs. (44) and (45),

$$h(\mathbf{r},t) = \frac{3}{2}p(\mathbf{r},t) + n(\mathbf{r},t)\frac{1}{2}mv^2(\mathbf{r},t), \quad (48)$$

$$\mathbf{I}(\mathbf{r},t) = n(\mathbf{r},t)\mathbf{v}(\mathbf{r},t) = -\beta^*(\mathbf{r},t)P^{[2]}(\mathbf{r},t)\cdot\mathbf{v}(\mathbf{r},t), \quad (49)$$

after using Eqs. (B14) and (B17), and finally [cf. Eq. (46)]

$$I_n^{[2]}(\mathbf{r},t) = \frac{1}{m}P^{[2]}(\mathbf{r},t) + n(\mathbf{r},t)m[\mathbf{v}(\mathbf{r},t)\mathbf{v}(\mathbf{r},t)]. \quad (50)$$

On the other hand, $n(\mathbf{r},t)$ is given in Eqs. (B11) and (B12) in Appendix B.

To proceed further, and in order to avoid cumbersome expressions that would obscure the analysis of the physics involved, we consider the case in which are null the components $P_{23}$, $P_{32}$, $P_{13}$, $P_{31}$, of the pressure tensor: this implies in a planar Couette flow. Using Eq. (46) and that $\mathcal{M}^{[2]}$ is the inverse of $M^{[2]}$ of Eq. (45), one finds that (in the expressions below we omit—for typographical simplicity—the dependence of the variables on position and time, i.e., we do not explicitly write $(\mathbf{r},t)$ which is kept implicit)

$$M_{11} = \beta^* + \frac{2}{m}F_{11} = \frac{nP_{22}}{\Delta}, \quad (51)$$

$$M_{22} = \beta^* + \frac{2}{m}F_{22} = \frac{nP_{11}}{\Delta}, \quad (52)$$

$$M_{33} = \beta^* + \frac{2}{m}F_{33} = \frac{nP_{33}}{\Delta}, \quad (53)$$

where

$$\Delta = P_{11}P_{22} - (P_{12})^2, \quad (54)$$

and we have used that

$$M = \det M^{[2]} = \frac{n^3}{P_{33}}\Delta. \quad (55)$$

These equations relate $\beta^*$ and $F_{ij}$, ($j = 1, 2, 3$), with the pressure tensor, and Eq. (44) relates $\beta^*$ and $\mathbf{v}$ with the energy density.

The denominator $\Delta$ [as given by Eq. (54)] in Eqs. (51)–(53) may indicate the possibility of a singularity and then of a hydrodynamic instability, but this is not the case, as shall be shown later on. Taking into account Eq. (46), that is, that the tensor $F^{[2]}$ is traceless, we solve Eqs. (51)–(55) to obtain that

$$\beta^* = \frac{1}{3}n\left[\frac{P_{11} + P_{22}}{\Delta} + \frac{1}{P_{33}}\right], \quad (56)$$

$$-\frac{2}{m}F_{11} = \frac{1}{3P_{33}}\frac{n}{\Delta}\left[P_{22}(P_{11} - P_{33}) + P_{33}(P_{11} - P_{22}) - (P_{12})^2\right], \quad (57)$$

$$-\frac{2}{m}F_{22} = \frac{1}{3P_{33}}\frac{n}{\Delta}\left[P_{11}(P_{22} - P_{33}) + P_{33}(P_{22} - P_{11}) - (P_{12})^2\right], \quad (58)$$

$$-\frac{2}{m}F_{33} = \frac{1}{3P_{33}}\frac{n}{\Delta}\left[P_{11}(P_{33} - P_{22}) + P_{22}(P_{11} - P_{33}) + 2(P_{12})^2\right], \quad (59)$$

$$-\frac{2}{m}F_{12} = \frac{n}{\Delta}P_{12}. \quad (60)$$

In the limit of equilibrium is follows that $F_{ij} = 0$, $\mathbf{v} = 0$, and $\beta^* \to [k_B T]^{-1}$ where $T$ is temperature in equilibrium as it should. Moreover, on inspection of Eq. (56) it can be noticed that we may write

$$\frac{1}{T^*} = \frac{1}{3}\left[\frac{1}{\Theta_{11}} + \frac{1}{\Theta_{22}} + \frac{1}{\Theta_{33}}\right], \quad (61)$$

introducing a characteristic nonequilibrium temperature (quasi-temperature or kinetic temperature[45]), for each of the three spatial directions,[46] namely,

$$nk_B\Theta_{11} = \frac{\Delta}{P_{11}}, \quad (62)$$

$$nk_B\Theta_{22} = \frac{\Delta}{P_{22}}, \quad (63)$$

$$nk_B\Theta_{33} = \frac{\Delta}{P_{33}}. \quad (64)$$

When $P_{12} = 0$, these three expressions coincide, $\Theta_{11} = \Theta_{22} = \Theta_{33}$, and it is verified an equipartition law in terms of $\Theta$, namely, $P_{11} = P_{22} = P_{33} = P_{kin} = nk_B\Theta$, where $P_{kin}$ is the kinetic pressure.

After the derivation we have performed in this section of the non-equilibrium equations of state, relating the basic thermo-hydrodynamic variables, we proceed to obtain their equations of motion in Sec. IV. Previously, we analyze, in this expanded (second-order) grand-canonical description the nonequilibrium Shannon entropy and entropy production densities, the quantity akin to a free energy density, and the quasi-chemical potential.

## III. QUASI-THERMODYNAMIC FUNCTIONS

We consider here the fields of nonequilibrium information-theoretic entropy, its production, and the quasi-free energy and quasi-chemical potential defined in the present description.

The nonequilibrium information-theoretic entropy, here [i.e., the average over the nonequilibrium ensemble of the quantity of Eqs. (5) and (38)], is

$$\bar{S}(t) = \int d^3 r\, \bar{s}(\mathbf{r}, t), \tag{65}$$

where we have introduced the quasi-entropy density,

$$\bar{s}(\mathbf{r}, t) = -\beta^*(\mathbf{r}, t)\Omega(\mathbf{r}, t) + \beta^*(\mathbf{r}, t) h(\mathbf{r}, t) \\ - \beta^*(\mathbf{r}, t)\mu^*(\mathbf{r}, t) n(\mathbf{r}, t) \\ - \mathbf{F}_n(\mathbf{r}, t) \cdot \mathbf{I}_n(\mathbf{r}, t) - F_n^{[2]}(\mathbf{r}, t) \otimes I_n^{[2]}(\mathbf{r}, t), \tag{66}$$

and we have written [cf. Eq. (39)]

$$\phi(t) = \ln \bar{Z}(t) = -\int d^3 r\, \beta^*(\mathbf{r}, t)\Omega(\mathbf{r}, t), \tag{67}$$

introducing the free energy density (characteristic of this description) $\Omega(\mathbf{r}, t)$. In terms of the pressure tensor and the barycentric velocity, we can write (we omit to explicitly write the dependence on $(\mathbf{r}, t)$ on the right side)

$$\bar{s}(\mathbf{r}, t) = -\beta^* \Omega + \beta^* \left(u + n\frac{1}{2}mv^2\right) - \left(\beta^*\mu^* + mF_n^{[2]} \otimes [\mathbf{vv}]\right) n \\ - \mathbf{F}_n \cdot \mathbf{I}_n - \frac{1}{m} F_n^{[2]} \otimes \mathring{P}^{[2]}], \tag{68}$$

where, we recall, $\mathrm{Tr}\{F_n^{[2]}\} = 0$ and then $F_n^{[2]} \otimes P^{[2]}$ with $\mathring{P}^{[2]}$ being the traceless part of the pressure tensor, and $u$ is the internal energy of Eq. (45). Using Eqs. (44), (49), and (50), after some algebra we find that

$$\bar{s}(\mathbf{r}, t) = \beta^* \Omega - \beta^* \mu^* - n\frac{1}{2}mM^{[2]} \otimes [\mathbf{vv}] + \frac{3}{2}n. \tag{69}$$

Taking into account that $\Omega = n/\beta^*$ [cf. Eq. (75)], and introducing the quantity,

$$\tilde{\mu}(\mathbf{r}, t) = \beta^*\mu^* + \frac{1}{2}mM^{[2]} \otimes [\mathbf{vv}] \\ \tilde{\mu}(\mathbf{r}, t) = \beta^*\mu^* + \beta^*\frac{1}{2}mv^2 - mF_n^{[2]} \otimes [\mathbf{vv}] \\ \tilde{\mu}(\mathbf{r}, t) = \beta^*\left(\mu^* + \frac{1}{2}mv^2 - \beta^* mv_n^{[2]} \otimes [\mathbf{vv}]\right), \tag{70}$$

we do have that

$$\bar{s}(\mathbf{r}, t) = n(\mathbf{r}, t)\left(\frac{5}{2} - \beta^*(\mathbf{r}, t)\tilde{\mu}(\mathbf{r}, t)\right). \tag{71}$$

It can be noticed that in the limit of equilibrium, it is recovered the well-known expression for the entropy in the grand-canonical ensemble.

For the quasi-entropy production, using Eqs. (65) and (66), it follows that

$$\bar{\sigma}(t) = \frac{d}{dt}\bar{S}(t) = \int d^3 r\, \frac{\partial}{\partial t}\bar{s}(\mathbf{r}, t),$$

$$\bar{\sigma}(t) = \int d^3 r\, \beta^*\left[\frac{\partial h}{\partial t} - \mu^*\frac{\partial n}{\partial t} + m\mathbf{v}_n \cdot \frac{\partial \mathbf{I}_n}{\partial t} + m^2 \beta^* v_n^{[2]} \otimes \frac{\partial I_n^{[2]}}{\partial t}\right], \tag{72}$$

once it is verified that

$$\frac{d\phi}{dt} + \int d^3 r\left[\frac{\partial \beta^*}{\partial t} h - nm\frac{\partial}{\partial t}(\beta^*\mu^*) + \frac{\partial \mathbf{F}_n}{\partial t} \cdot \mathbf{I}_n - \frac{\partial F_n^{[2]}}{\partial t} \otimes I_n^{[2]}\right] = 0, \tag{73}$$

which is a kind of Gibbs–Duhem relation[47] in these nonequilibrium conditions.

It can be noticed that this quasi-entropy production function has not a definite sign. It is composed of two parts: $\bar{\sigma}_i(t) + \bar{\sigma}_e(t)$, consisting of the one associated with internal relaxation process, and a second one associated with interactions with the surroundings (in the steady state there follows a net balance, resulting in a zero global entropy variation).[2,3,48] We can see in Eq. (72) that the quasi-entropy production function is determined by the equations of evolution of the basic variables, which are given in Sec. IV. In those equations, we can separate in the collision integrals the part due to internal interactions and the one due to the interaction with external sources, which, respectively, are the sources of the internal and external contributions to the quasi-entropy productions. It can also be noticed that, even though the quasi-entropy increases with respect to its initial value, it is not a monotonic increase: the internal quasi-entropy production $\sigma_i(t)$ may have a negative derivative in certain intervals. This is a result of the use of contracted descriptions as done here, what is discussed in Refs. 2,3,49, and 50.

The quasi-free energy satisfies that

$$\ln \bar{Z}(t) = \int d^3 r\, \beta^*(\mathbf{r}, t)\Omega(\mathbf{r}, t),$$

$$\ln \bar{Z}(t) = \left(\frac{2\pi m}{\hbar^2}\right)^{3/2} \int d^3 r\, M^{-1/2} \times \exp\left\{\beta^*\mu^* + \frac{1}{2}mM^{[2]} \otimes [\mathbf{vv}]\right\},$$

$$\ln \bar{Z}(t) = \left(\frac{2\pi m}{\hbar^2}\right)^{3/2} \int d^3 r\, M^{-1/2} e^{\beta^*\tilde{\mu}}. \tag{74}$$

The expression for $\Omega(\mathbf{r}, t)$ follows the identifying both quantities under the integral sign, and taking into account Eq. (B11), we do have that

$$\Omega(\mathbf{r}, t) = -\frac{n(\mathbf{r}, t)}{\beta^*(\mathbf{r}, t)} \equiv p_{kin}(\mathbf{r}, t), \tag{75}$$

that is, minus what can be called the nonequilibrium kinetic pressure. In the limit of equilibrium, there follows the known expression for the grand-canonical free energy, and the usual equation of state $pV = Nk_B T$ for the ideal gas.

For the nonequilibrium chemical potential (quasi-chemical potential) in the description in terms of the set of variables in Eq. (19), we obtain, using the generalization of the usual form in equilibrium, that

$$\beta^*(\mathbf{r},t)\mu^*(\mathbf{r},t) = \frac{\delta\bar{\sigma}(\mathbf{r},t)}{\delta n(\mathbf{r},t)}, \qquad (76)$$

after using Eq. (66).

On the other hand using the representation which introduces $\mathbf{v}$ and the pressure tensor, it is taken as the $\tilde{\mu}(\mathbf{r},t)$ of Eq. (70). Using Eq. (70), it follows that [cf. Eq. (B11)]

$$n(\mathbf{r},t) = \left(\frac{2\pi m}{\hbar^2}\right)^{3/2} \frac{e^{\beta^*(\mathbf{r},t)\tilde{\mu}(\mathbf{r},t)}}{\sqrt{M(\mathbf{r},t)}}, \qquad (77)$$

and then

$$\mu^*(\mathbf{r},t) = -\frac{3}{2}k_B T^*(\mathbf{r},t) \ln\left\{\frac{[M(\mathbf{r},t)]^{-1/3}}{k_B \theta_k(\mathbf{r},t)}\right\}, \qquad (78)$$

where

$$k_B \theta_k(\mathbf{r},t) = \frac{\hbar^2 [n(\mathbf{r},t)]^{2/3}}{2\pi m},$$

and in the limit of null equilibrium is recovered the usual result,

$$\mu = -\frac{3}{2}k_B T \ln(T/\theta_k),$$

where $\hbar^2[n(\mathbf{r},t)]^{2/3}/2\pi m$ is the so-called characteristic temperature associated with the kinetic motion.

## IV. THE KINETIC EQUATIONS

Calling generically as $Q_j(\mathbf{r},t)$ the set of basic variables of Eqs. (13)–(18), and $P_j(\mathbf{r},t)$ the corresponding dynamical quantities, their equations of motion are

$$\frac{\partial}{\partial t}Q_j(\mathbf{r},t) = J_j^{(0)}(\mathbf{r},t) + J_j^{(1)}(\mathbf{r},t) + \mathcal{J}_j(\mathbf{r},t) + \mathcal{F}_j(\mathbf{r},t), \qquad (79)$$

where

$$J_j^{(0)}(\mathbf{r},t) = \int d\Gamma \int d\Gamma_B \{\hat{P}_j(\mathbf{r}), \hat{H}_0\} \bar{\varrho}(t,0) \times \rho_B, \qquad (80)$$

$$J_j^{(1)}(\mathbf{r},t) = \int d\Gamma \int d\Gamma_B \{\hat{P}_j(\mathbf{r}), \hat{H}'\} \bar{\varrho}(t,0) \times \rho_B, \qquad (81)$$

$$\mathcal{J}_j(\mathbf{r},t) = \int d\Gamma \int d\Gamma_B \{\hat{P}_j(\mathbf{r}), \hat{H}'\} \varrho'_\varepsilon(t,0) \times \rho_B, \qquad (82)$$

and $\mathcal{F}_j(\mathbf{r},t)$ accounts for the effect of the external forces (presence of $V_{ext}$), which is responsible for inducing the flow given by

$$\mathcal{F}_j(\mathbf{r},t) = \int d\Gamma \int d\Gamma_B \{\hat{P}_j(\mathbf{r}), V_{ext}\} \varrho_\varepsilon(t,0) \times \rho_B. \qquad (83)$$

These kinetic equations are the classical-mechanics' Hamilton equations for quantities $\hat{P}_j(\mathbf{r})$ averaged over the nonequilibrium ensemble, and where

$$\hat{H}_0 = \int d^3r \int d^3p \, \frac{p^2}{2m} \hat{n}_1(\mathbf{r},\mathbf{p}), \qquad (84)$$

$$\hat{H}' = \sum_{i,j} \vartheta(|\mathbf{r}_j - \mathbf{R}_i|), \qquad (85)$$

and use was made of separation of the statistical operator of Eq. (3) in the form

$$\varrho_\varepsilon(t) = \bar{\varrho}(t,0) + \varrho'_\varepsilon(t), \qquad (86)$$

with $\bar{\varrho}(t,0)$ being the auxiliary operator of Eq. (20), that is, the "instantaneously frozen quasi-equilibrium" statistical operator, and $\varrho'_\varepsilon(t)$ is the contribution responsible for including relaxation processes (combining historicity and irreversibility), as described in Refs. 10 and 25. Moreover, $\rho_B$ is the canonical distribution of the thermal bath in equilibrium at temperature $T_0$, with $d\Gamma_B$ being the corresponding infinitesimal volume element in phase space, while $d\Gamma$ is given in Eq. (23).

The collision integrals $\mathcal{J}_j(\mathbf{r},t)$ can be handled in a perturbational-like approach, when it takes the form of a series of partial collision integrals,

$$J_j(\mathbf{r},t) = \sum_{l=2}^{\infty} \Omega_j^{(l)}(\mathbf{r},t), \qquad (87)$$

and the $\Omega_j$ can also be decomposed in the form

$$\Omega_j(\mathbf{r},t) = \sum_{n \geqslant l}^{\infty} J_{(l)}^{(n)}(\mathbf{r},t), \qquad (88)$$

in terms of "instantaneously frozen" correlations functions encompassing the auxiliary operator $\bar{\varrho}(t,0)$ at time.[10–12,27,28]

These partial collision integrals are, in general, composed of the average over the auxiliary ensemble of Born's perturbation series (involving binary, tertiary, etc., collisional processes), but accompanied by a kind of "vortex renormalization" and "propagator renormalization" in and between collisional events.[27] Keeping only—what we do in what follows—the first contribution $J^{(2)}$ alone is to resort to a Markonian approximation. Moreover, $J^{(1)}$ is null in all the four equations, $J^{(0)}$ gives rise to the part of the constitutive equations involving the divergence of a flux, and $J^{(2)}$ is reduced to the Golden Rule of perturbation theory averaged over the "instantaneously frozen" nonequilibrium macrostate.[10,27,28]

Now, we restrict the calculation to the use of a nonlinear higher-order hydrodynamics of order 2 (see Appendix C), that is to report to a contracted description consisting in to retain as basic dynamical variables the set

$$\left\{h(\mathbf{r},t), n(\mathbf{r},t), \mathbf{I}_n(\mathbf{r},t), I_n^{[2]}(\mathbf{r},t)\right\}.$$

The corresponding equations of evolution for the last three are

$$\frac{\partial}{\partial t}n(\mathbf{r},t) + \nabla \cdot \mathbf{I}_n(\mathbf{r},t) = 0, \qquad (89)$$

$$\frac{\partial}{\partial t}\mathbf{I}_n(\mathbf{r},t) + \nabla \cdot I_n^{[2]}(\mathbf{r},t) = \mathbf{J}_n(\mathbf{r},t) + F(\mathbf{r},t), \qquad (90)$$

$$\frac{\partial}{\partial t}I_n^{[2]}(\mathbf{r},t) + \nabla \cdot I_n^{[3]}(\mathbf{r},t) = J_n^{[2]}(\mathbf{r},t) + F^{[2]}(\mathbf{r},t), \qquad (91)$$

where

$$F^{[2]}(\mathbf{r},t) = [\mathbf{v}(\mathbf{r},t)F(\mathbf{r},t)], \tag{92}$$

and $\mathbf{J}_n(\mathbf{r},t)$ and $J_n^{[2]}$ have complicate expressions—see Appendix C—which are dependent on position and changing in time because they depend on the nonequilibrium thermodynamic state of the system. As noticed in Appendix C, those collision integrals are composed of three contributions, of which the main one is the form

$$\mathbf{J}_n(\mathbf{r},t) = -A^{[2]}(\mathbf{r},t)\mathbf{I}_n(\mathbf{r},t),$$

for the first flux, and

$$J_n^{[2]}(\mathbf{r},t) = -A^{[3]}(\mathbf{r},t)I_n^{[2]}(\mathbf{r},t),$$

where $A^{[2]}$ and $A^{[3]}$ have dimensions of inverse of time, playing the role of inverse of tensorial Maxwell's characteristic times.

However, for the case of Brownian particles, that we are considering for illustration, they take the simple forms,

$$\mathbf{J}_n(\mathbf{r},t) = -\tau_{n_1}^{-1}\mathbf{I}_n(\mathbf{r},t), \tag{93}$$

$$\mathbf{J}_n^{[2]}(\mathbf{r},t) = \frac{n(\mathbf{r},t)}{m\beta_B}\tau_{n_2}^{-1}\mathbf{1}^{[2]} - \tau_{n_2}^{-1}\mathbf{I}_n^{[2]}(\mathbf{r},t), \tag{94}$$

with $\tau_{n_1}$ and $\tau_{n_2}$ playing the role of Maxwell's characteristic times associated with the flux and the second-order flux;[51,52] see in Appendix C the conditions for obtaining this simple forms for the collision integrals, and the expression for them.

Taking into account that $\mathbf{I}(\mathbf{r},t) = n(\mathbf{r},t)\mathbf{v}(\mathbf{r},t)$ [cf. Eq. (B14)], where $\mathbf{v}(\mathbf{r},t)$ is the barycentric velocity, and Eq. (50) which introduces the pressure tensor, from Eqs. (89)–(91)—which are the evolution equations for the quantities of set (19)—can be derived the equations of evolution for the set

$$\left\{h(\mathbf{r},t), n(\mathbf{r},t), \mathbf{v}(\mathbf{r},t), P^{[2]}(\mathbf{r},t)\right\}. \tag{95}$$

They are

$$\frac{\partial}{\partial t}n(\mathbf{r},t) = -\nabla \cdot [n(\mathbf{r},t)\mathbf{v}(\mathbf{r},t)], \tag{96}$$

$$n(\mathbf{r},t)\frac{\partial}{\partial t}\mathbf{v}(\mathbf{r},t) = -\frac{1}{m}\nabla \cdot P^{[2]}(\mathbf{r},t) - n(\mathbf{r},t)[\mathbf{v}(\mathbf{r},t) \cdot \nabla]\mathbf{v}(\mathbf{r},t)$$
$$-\tau_{n_1}^{-1}n(\mathbf{r},t)\mathbf{v}(\mathbf{r},t) + \mathbf{F}(\mathbf{r},t), \tag{97}$$

$$\frac{\partial}{\partial t}P_n^{[2]}(\mathbf{r},t) = -\nabla \cdot I_n^{[3]}(\mathbf{r},t) - m\frac{\partial}{\partial t}(n(\mathbf{r},t)[\mathbf{v}(\mathbf{r},t)\mathbf{v}(\mathbf{r},t)])$$
$$-\frac{n(\mathbf{r},t)}{m\beta_B}\tau_{n_2}^{-1}\mathbf{1}^{[2]} - \tau_{n_2}^{-1}P^{[2]}(\mathbf{r},t) + \frac{1}{m}[\mathbf{F}(\mathbf{r},t)\mathbf{v}(\mathbf{r},t)]. \tag{98}$$

Moreover, we can see that in Eq. (98) it is present the third-order flux $I^{[3]}(\mathbf{r},t)$ [cf. Eq. (17)], which is not a basic variable and then to close the system of equations is necessary to express it in terms of the basic variables, what is done to obtain that

$$m\nabla \cdot I_n^{[3]}(\mathbf{r},t) = (\nabla \cdot \mathbf{v}(\mathbf{r},t))P^{[2]}(\mathbf{r},t) + P^{[2]}(\mathbf{r},t) \cdot V^{[2]}(\mathbf{r},t)$$
$$+ (V^{[2]}(\mathbf{r},t))^{tr} \cdot P^{[2]}(\mathbf{r},t) + (\mathbf{v}(\mathbf{r},t) \cdot \nabla)P^{[2]}(\mathbf{r},t)$$
$$+ [\nabla \cdot P^{[2]}(\mathbf{r},t)\mathbf{v}(\mathbf{r},t)] + [\mathbf{v}(\mathbf{r},t)\nabla \cdot P^{[2]}(\mathbf{r},t)]$$
$$+ m\nabla \cdot [n(\mathbf{r},t)\mathbf{v}(\mathbf{r},t)\mathbf{v}(\mathbf{r},t)\mathbf{v}(\mathbf{r},t)], \tag{99}$$

after using Eq. (50) for expressing it in terms of the pressure tensor, and where

$$V^{[2]}(\mathbf{r},t) = [\nabla\mathbf{v}(\mathbf{r},t)] \tag{100}$$

and [tr] stands for transposed (we recall that [···] stands for tensorial product of vectors).

From the set of Eqs. (89)–(91), deriving twice in time Eq. (89), deriving once Eq. (90) and next using in it Eq. (91), the results can be combined to obtain the following equation:

$$\left(\frac{\partial^3}{\partial t^3} + \tau_{n+}^{-1}\frac{\partial^2}{\partial t^2} + \tau_{nx}^{-1}\frac{\partial}{\partial t}\right)n(\mathbf{r},t) - \tau_{n_2}^{-1}\nabla^2\left(\frac{n(\mathbf{r},t)}{m\beta_B}\right)$$
$$+ \nabla \cdot \nabla \cdot \nabla \cdot I_n^{[3]}(\mathbf{r},t) = \Phi_{ext}(\mathbf{r},t), \tag{101}$$

where

$$\Phi_{ext}(\mathbf{r},t) = \frac{1}{m}\nabla \cdot \nabla \cdot F_n^{[2]}(\mathbf{r},t) - \frac{\tau_{n_2}^{-1}}{m}\nabla \cdot \mathbf{F}(\mathbf{r},t)$$
$$- \frac{1}{m}\frac{\partial}{\partial t}\nabla \cdot \mathbf{F}(\mathbf{r},t) \tag{102}$$

is a source term,

$$\tau_{n+}^{-1} = \tau_{n_1}^{-1} + \tau_{n_2}^{-1}, \tag{103}$$

$$\tau_{nx}^{-1} = \tau_{n_1}^{-1}\tau_{n_2}^{-1}, \tag{104}$$

and we recall the divergence of the third-order flux is given in Eq. (99). Equation (101) is then coupled to the equations of evolution for the pressure tensor and for the velocity. Moreover, we can see that it is a third-order equation in time, which is characteristic of this second-order approximation in the fluxes (see below); neglecting the third-order derivative in time (movements with small change in time), there follows a Maxwell–Cattaneo-like (hyperbolic type) equation, and further neglecting the second derivative in time (even smaller change in time) one recovers a Fick-like (parabolic-type) equation of motion.

On this, we recall that an equation of motion for the density of n-degree in the time derivative results in a description that includes up to the $(n-1)$-order flux in the set of basic variables. Thus, in a zeroth-order description there follows an equation of motion which is first-degree in the time derivative (diffusion equation); a first-order description leads to an equation of motion of second-degree in the time derivative (Maxwell–Cattaneo), and so forth. The criterion that establishes the order of the truncation for having a satisfactory description of the movement depends on the characteristics of this movement: in a qualitative way, we can say that a lower and lower order of description is acceptable accompanying movements in which the relevant contributions are, correspondingly, of larger and larger wavelengths (smaller and smaller wavenumbers) and with accompanying smaller and smaller frequencies (as known, a diffusion equation is satisfactory in near static and near uniform conditions). On the other hand, on a quantitative level, criteria can be established,[32] and we apply them to the present case.

## A. Hyperbolic equation with advection and viscous pressure

A Maxwell–Cattaneo-like equation when considering the second-order flux as nearly time independent and then from Eqs. (91) and (94), we do find that

$$\nabla \cdot I_n^{[3]}(\mathbf{r}, t) = \frac{n(\mathbf{r}, t) \tau_{n_2}^{-1}}{m \beta_B} - \tau_{n_2}^{-1} I^{[2]}(\mathbf{r}, t) + \tau_{n_2}^{-1} F^{[2]}(\mathbf{r}, t), \quad (105)$$

and then Eq. (97) takes the form

$$\left(\tau_{n+}^{-1} \frac{\partial^2}{\partial t^2} + \tau_{nx}^{-1} \frac{\partial}{\partial t}\right) n(\mathbf{r}, t) - \tau_{n_2}^{-1} \nabla \cdot \nabla \cdot I_n^{[2]}(\mathbf{r}, t)$$
$$= -\tau_{n_2}^{-1} \nabla \cdot \mathbf{F}(\mathbf{r}, t) - \frac{\partial}{\partial t} \nabla \cdot \mathbf{F}(\mathbf{r}, t). \quad (106)$$

For convenience, we write

$$\tau_{n_1} \mathbf{F}(\mathbf{r}, t) \equiv n(\mathbf{r}, t) \mathbf{u}_{\text{ext}}(\mathbf{r}, t), \quad (107)$$

defining the quantity $\mathbf{u}_{\text{ext}}$ which has dimensions of velocity, and the right of Eq. (106) can be interpreted as a result of the presence of the driven flux forced by the application of the external force. Using that

$$I_n^{[2]}(\mathbf{r}, t) = \frac{1}{m} P^{[2]}(\mathbf{r}, t) + n(\mathbf{r}, t)[\mathbf{v}(\mathbf{r}, t)\mathbf{v}(\mathbf{r}, t)], \quad (108)$$

after multiplying by $\tau_{nx}^{-1}$, Eq. (106) becomes

$$\left[(\tau_{n_1} + \tau_{n_2}) \frac{\partial^2}{\partial t^2} + \frac{\partial}{\partial t}\right] n(\mathbf{r}, t) - \frac{\tau_{n_1}}{m} \nabla \cdot \nabla \cdot \tilde{P}^{[2]}(\mathbf{r}, t) = G_n(\mathbf{r}, t), \quad (109)$$

where

$$\tilde{P}^{[2]}(\mathbf{r}, t) = P^{[2]}(\mathbf{r}, t) + mn(\mathbf{r}, t)[\mathbf{v}(\mathbf{r}, t)\mathbf{v}(\mathbf{r}, t)], \quad (110)$$

which is $mI_n^{[2]}(\mathbf{r}, t)$, cf. Eq. (108), and

$$G_n(\mathbf{r}, t) = -\nabla \cdot (n(\mathbf{r}, t)\mathbf{u}_{\text{ext}}(\mathbf{r}, t)) - \tau_{n_2} \frac{\partial}{\partial t} \nabla \cdot (n(\mathbf{r}, t)\mathbf{u}_{\text{ext}}(\mathbf{r}, t)), \quad (111)$$

with contributions associated with the external force.

We write the pressure tensor in the form

$$P^{[2]}(\mathbf{r}, t) = \frac{n(\mathbf{r}, t)}{\beta^*(\mathbf{r}, t)} 1^{[2]} P_v^{[2]}(\mathbf{r}, t), \quad (112)$$

consisting of a kinetic contribution [cf. Eq. (75)] plus a viscous contribution (the later contains the nondiagonal shear contributions depending on the shear rate on which also depends the remaining diagonal parts, as we shall see as we proceed). Hence, using this Eq. (112), Eq. (109) becomes

$$\left[(\tau_{n_1} + \tau_{n_2}) \frac{\partial^2}{\partial t^2} + \frac{\partial}{\partial t}\right] n(\mathbf{r}, t) - \nabla^2 \left[D_n^{[2]}(\mathbf{r}, t) n(\mathbf{r}, t)\right]$$
$$= \frac{\tau_{n_1}}{m} \nabla \cdot \nabla \cdot P_v^{[2]}(\mathbf{r}, t) + G_n(\mathbf{r}, t), \quad (113)$$

where

$$D_n^{[2]}(\mathbf{r}, t) = \frac{\tau_{n_1}}{m} \left[k_B T^*(\mathbf{r}, t) 1^{[2]} + m[\mathbf{v}(\mathbf{r}, t)\mathbf{v}(\mathbf{r}, t)]\right], \quad (114)$$

and we recall that we have written $k_B T^*(\mathbf{r}, t) = 1/\beta^*(\mathbf{r}, t)$. $D$ plays the role of a space and time dependent diffusion coefficient, which is composed of two contributions, one associated with the thermal energy (thermal motion) and the other to the kinetic energy in the drift motion.

Equation (113) has the form of a Maxwell–Cattaneo hyperbolic equation, with a rigid-hand side with sources. The latter are of two kinds: one contribution arises out of the external perturbation (the term containing $u_{\text{ext}}$ in $G_n$), another (quadratic in the velocity) associated with the convective pressure, and a third one containing the effect of the viscous pressure.

We emphasize that, for the equation for viscous pressure, a relaxational equation analogous to Maxwell–Cattaneo one is often dubbed as "Maxwell viscoelastic equation."

### B. Diffusion-advection equation

A Fick-like equation follows when the first-order flux is taken as nearly time-independent; then from Eqs. (94), (101), and (112), we do have that

$$\nabla \cdot I_n^{[2]}(\mathbf{r}, t) = -\tau_{n_1}^{-1} I_n(\mathbf{r}, t) + \tau_{n_1}^{-1} n(\mathbf{r}, t) u_{\text{ext}}(\mathbf{r}, t). \quad (115)$$

Replacing in this equation the second flux in terms of the pressure tensor and the convective pressure as indicated in Eq. (42), and after introducing the expression for $I_n$, obtained from Eq. (115) in Eq. (89), it follows that

$$\frac{\partial}{\partial t} n(\mathbf{r}, t) - \frac{\tau_{n_1}}{m} \nabla \cdot \nabla \cdot \tilde{P}^{[2]}(\mathbf{r}, t) + \nabla \cdot [n(\mathbf{r}, t) u_{\text{ext}}(\mathbf{r}, t)] = 0. \quad (116)$$

Performing the separation of the pressure tensor as indicated in Eq. (112), it follows that

$$\frac{\partial}{\partial t} n(\mathbf{r}, t) - \nabla \cdot \nabla \cdot \left[D_{n_1}^{[2]}(\mathbf{r}, t) n(\mathbf{r}, t)\right]$$
$$= \frac{\tau_{n_1}}{m} \nabla \cdot \nabla \cdot P_v^{[2]}(\mathbf{r}, t) - \nabla \cdot (n(\mathbf{r}, t) u_{\text{ext}}(\mathbf{r}, t)), \quad (117)$$

where $D_{n_1}^{[2]}(\mathbf{r}, t)$ is the one of Eq. (114), which, as noticed, plays the role of a diffusion coefficient varying in position and changing in time.

We can see that Eq. (117) can be written as an equation of conservation of the type,

$$\frac{\partial}{\partial t} n(\mathbf{r}, t) + \nabla \cdot \mathbf{I}^*(\mathbf{r}, t) = 0, \quad (118)$$

where

$$\mathbf{I}^*(\mathbf{r}, t) = -\nabla \cdot \left[D_{n_1}^{[2]}(\mathbf{r}, t) n(\mathbf{r}, t)\right] - \frac{\tau_{n_1}}{m} \nabla \cdot P_v^{[2]}(\mathbf{r}, t)$$
$$- n(\mathbf{r}, t) u_{\text{ext}}(\mathbf{r}, t) \quad (119)$$

is a flux composed of, respectively, a generalized diffusive contribution, the contribution from the viscous stress and, finally, the contribution associated with the flow generated by the external force.

For illustration, let us consider a one-dimensional system under driven flow. In this case, Eq. (117) becomes

$$\frac{\partial}{\partial t} n(\mathbf{r}, t) + \frac{\partial}{\partial x} \left[\frac{\partial}{\partial x} [D(x, t) n(x, t)] + n(x, t) u_{\text{ext}}(x)\right] = 0, \quad (120)$$

situation where the kinetic energy of drift can be neglected in comparison with the thermal energy $k_B T^*$, that is a very good thermal contact with the thermal bath, and that $T^* \simeq T_B$, and then $D \simeq (\tau_{n_1}/m) k_B T_B$, a constant in time and space. Moreover, we take

$$u_{\text{ext}}(x) = k_1 - k_2 x, \quad (121)$$

composed of a contribution leading to streaming and other associated with a harmonic force.[53] It could be the case of an electrolyte—the Brownian particles are charged—in the presence of an electric field in $x$ direction and increasing linearly along the $x$ axis, or forcing such movement with optical tweezers. Thus, the contribution of Eq. (120) is

$$n(x,t) = \xi^{-1}(t) e^{-b(t)[x-\bar{x}(t)]^2}, \quad (122)$$

where

$$\bar{x}(t) = \frac{k_1}{k_2} + \left(\bar{x}(0) - \frac{k_1}{k_2}\right) e^{-k_2 t}, \quad (123)$$

$$\frac{b(t)}{b(0)} = \left(\frac{\xi(0)}{\xi(t)}\right)^2 \frac{1}{(1-\Delta t)e^{-2k_2 t} + \Delta t}, \quad (124)$$

$$\Delta t = \frac{2Db(0)}{k_2}, \quad (125)$$

and $b(0)$ and $\xi(0)$ follow from the constraints:

$$m_0 L - \frac{1}{n_0 L}\int dx\, n(x,t) = \frac{\sqrt{\pi}}{\xi(0)}\sqrt{b(0)}, \quad (126)$$

$$\int_0^L dx(x-\bar{x}(t))^2 n(x,t) = \frac{n_0 L}{2b(t)}, \quad (127)$$

where $n_0$ is the whole concentration and $L$ the extent of the sample.

We further particularized the solution by taking $k_1 = 0$ and $k_2 = 1/\tau_0$, i.e., $u_{\text{ext}} = Vx/\tau_0$, where $\tau_0$ is a constant with dimensions of time. This corresponds to use Ornstein–Uhlenbeck model.[53] Once the driving force is time independent, the density, after a transient period, attains a stationary state, $n(x)$, satisfying the equation

$$\frac{d}{dx}\left[\frac{d}{dx}[D(x)n(x)] + \frac{x}{\tau_0}n(x)\right] \equiv \frac{d}{dx}I_n^*(x) = 0, \quad (128)$$

and then the flux $I_n^*$ (diffusive plus forced) is a constant, say, $I_0^*$, and then

$$\frac{d}{dx}[D(x)n(x)] + \frac{x}{\tau_0}n(x) = I_0^*. \quad (129)$$

Using the method of the integrating factors, it can be obtained the solution of Eq. (129), which is

$$n(x) = n(0)e^{-\psi(x)} + I_0^* e^{-\psi(x)} \int_0^x dx' \frac{e^{\psi(x')}}{D(x')}, \quad (130)$$

where

$$\psi(x) = \int_0^x dx' \frac{x'}{\tau_0} D(x'). \quad (131)$$

Moreover, we recall that it should be verified the constraint

$$n_0 = \int_0^L dx\, n(x), \quad (132)$$

which fixes the value of $n(0)$, and in the conditions stated above, the diffusion coefficient is a constant, say, $D$. Thus, Eq. (130) becomes

$$n(x) = n(0)e^{-x^2/D\tau_0} + \frac{I_0^*}{D} e^{-x^2/D\tau_0} \int_0^x dx'\, e^{-x'^2/D\tau_0},$$

$$n(x) = e^{-x^2/D\tau_0}\left[n(0) + \frac{1}{2}I_0^* \tau_0 (e^{x^2/D\tau_0} - 1)\right]. \quad (133)$$

## C. Evolution equations for the velocity, the pressure tensor, and the energy

The evolution equation for the velocity field is Eq. (97), which, after using Eq. (107) for the external force, becomes

$$n\frac{\partial \mathbf{v}}{\partial t} + n(\mathbf{v}\cdot\nabla)\mathbf{v} + \frac{1}{m}\nabla\cdot P^{[2]} = -n\frac{\mathbf{v} - \mathbf{u}_{\text{ext}}(x)}{\tau_{n_1}}. \quad (134)$$

On the other hand, the evolution equation for the pressure tensor field is Eq. (98), and after using Eq. (107) for the external force, it becomes

$$\frac{\partial}{\partial t}P^{[2]} + \frac{\partial}{\partial t}(mn[\mathbf{vv}]) - \nabla\cdot I^{[3]}$$
$$= -\tau_{n_2}^{-1}(P^{[2]} + mn[\mathbf{vv}]) + \frac{\tau_{n_2}^{-1}}{m\beta_R}n\mathbf{1}^{[2]} + \frac{\tau_{n_1}^{-1}}{m}nu_{\text{ext}}\mathbf{v}, \quad (135)$$

with the divergence of the third-order flux, $I^{[3]}$, is given in Eq. (99).

We do have now a set of coupled evolution equations for the density, the velocity, and the pressure tensor $\{n(\mathbf{r},t), \mathbf{v}(\mathbf{r},t), P^{[2]}(\mathbf{r},t)\}$, namely, Eqs. (89), (134), and (135), respectively, together with the auxiliary Eq. (99). It can be noticed that deriving in time Eq. (134) and using Eq. (135), we obtain for the velocity field the cumbersome second-order in time evolution equation,

$$n\frac{\partial^2 v}{\partial t^2} + n\frac{\partial n}{\partial t}\frac{\partial \mathbf{v}}{\partial t} + \frac{\partial}{\partial t}(n(\mathbf{v}\cdot\nabla)\mathbf{v})$$
$$- \frac{1}{m}\nabla\cdot\left[\frac{\partial}{\partial t}(mn[\mathbf{vv}]) - \nabla\cdot I^{[3]} + \tau_{n_2}^{-1}\left(P^{[2]} + mn[\mathbf{vv}]\right)\right.$$
$$\left. - \frac{n\tau_{n_2}^{-1}}{m\beta_R}\mathbf{1}^{[2]} + \frac{n\tau_{n_1}^{-1}}{m}u_{\text{ext}}\mathbf{v}\right] = -n\frac{\mathbf{v} - \mathbf{u}_{\text{ext}}(x)}{\tau_{n_1}}, \quad (136)$$

and once again recalling that $\nabla\cdot I^{[3]}$ is presented in Eq. (99).

First, we write

$$P^{[2]}(\mathbf{r},t) = p(\mathbf{r},t)\mathbf{1}^{[2]} + \overset{\circ}{P}{}^{[2]}(\mathbf{r},t), \quad (137)$$

where

$$p(\mathbf{r},t) = \frac{1}{3}\text{Tr}\left\{P^{[2]}(\mathbf{r},t)\right\} \quad (138)$$

and $\overset{\circ}{P}{}^{[2]}$ is the traceless part of $P^{[2]}$. Introducing these definitions in Eq. (99), we find that

$$m\nabla\cdot I^{[3]} = p\nabla\cdot \mathbf{v}\mathbf{1}^{[2]} + 2p\Lambda^{[2]} + (\mathbf{v}\cdot\nabla)p\mathbf{1}^{[2]} + [\mathbf{v}\nabla p]$$
$$+ [\nabla p\mathbf{v}] + \nabla\cdot(n[\mathbf{vvv}]) + m\nabla\cdot I_n^{[3]}|_0, \quad (139)$$

where

$$\nabla \cdot I_n^{[3]}|_0 = (\nabla \cdot \mathbf{v}) \overset{\circ}{P}{}^{[2]} + \overset{\circ}{P}{}^{[2]} \cdot \nabla \mathbf{v} + [\nabla \mathbf{v}]^{tr} \cdot \overset{\circ}{P}{}^{[2]}$$
$$+ (\mathbf{v} \cdot \nabla) \overset{\circ}{P}{}^{[2]} + \nabla \cdot \overset{\circ}{P}{}^{[2]} + \mathbf{v} \nabla \cdot \overset{\circ}{P}{}^{[2]}, \quad (140)$$

and we recall

$$\Lambda_{ij} = \frac{1}{2}\left(\frac{\partial v_i}{\partial x_j} + \frac{\partial v_j}{\partial x_i}\right). \quad (141)$$

In order to analyze these equations, we shall introduce some restrictive conditions.

Next, considering the case

$$\omega \tau_{n_2} \ll 1,$$

we can take

$$\frac{\partial I_n^{[2]}}{\partial t} \simeq 0$$

in Eq. (91) and then using Eqs. (108) and (94), it follows that

$$P^{[2]} \simeq \frac{n}{\beta_R} \mathbf{1}^{[2]} - \tau_{n_2} m \nabla \cdot I_n^{[3]} - mn[\mathbf{vv}]. \quad (142)$$

We neglect the last term in this Eq. (137), i.e., the convective presence, and we assume a smooth spatial variation in $I^{[3]}$, i.e., $I_n^{[3]}|_0 \simeq 0$. In such conditions, we obtain that

$$m \nabla \cdot I_n^{[3]} \simeq p\left[\frac{5}{3}\nabla \cdot \mathbf{v}\mathbf{1}^{[2]} + 2\left(\Lambda^{[2]} - \frac{1}{3}\nabla \cdot \mathbf{v}\mathbf{1}^{[2]}\right)\right], \quad (143)$$

and then

$$P^{[2]} = \frac{n}{\beta_R}\mathbf{1}^{[2]} - \varsigma \nabla \cdot \mathbf{v}\mathbf{1}^{[2]} - 2\eta\left(\Lambda^{[2]} - \frac{1}{3}\nabla \cdot \mathbf{v}\mathbf{1}^{[2]}\right), \quad (144)$$

where

$$\varsigma(\mathbf{r}, t) = \frac{5}{3}\tau_{n_2}p(\mathbf{r}, t), \quad (145)$$

$$\eta(\mathbf{r}, t) = \tau_{n_2}p(\mathbf{r}, t). \quad (146)$$

Taking the divergence in Eq. (144) and introducing the result in Eq. (134), it follows that

$$n(\mathbf{r}, t)\left(\frac{\partial}{\partial t} + \mathbf{v}(\mathbf{r}, t) \cdot \nabla\right)\mathbf{v}(\mathbf{r}, t) + \frac{1}{m}\nabla p(\mathbf{r}, t)$$
$$= \frac{1}{m}\eta(\mathbf{r}, t)\nabla^2 \mathbf{v}(\mathbf{r}, t) + \frac{1}{m}\left[\varsigma(\mathbf{r}, t) + \frac{1}{3}\eta(\mathbf{r}, t)\right]\nabla[\nabla \cdot \mathbf{v}(\mathbf{r}, t)]$$
$$- \tau_{n_1} n(\mathbf{r}, t)[\mathbf{v}(\mathbf{r}, t) - \mathbf{u}_{ext}(\mathbf{r}, t)]. \quad (147)$$

This Eq. (147) has the form of a Navier–Stokes equal but accompanied of advection and relaxation effects. Moreover, $\varsigma(\mathbf{r}, t)$ plays the role of space and time-dependent (i.e., following the evolution of the nonequilibrium thermodynamic state of the custom) bulk and shear viscosity coefficient.

On the other hand, it can be derived a generalized Maxwell equation for the traceless part of the pressure tensor field. In the restricted conditions presently used, it follows that

$$\nabla \cdot I_n^{[3]} = \frac{1}{\beta_R}\left[\nabla \cdot \mathbf{I}_n^{[2]}\mathbf{1}^{[2]} + \nabla \mathbf{I}_n + (\nabla \mathbf{I}_n)^{tr}\right], \quad (148)$$

and after using the equation of continuity and the definition of $\overset{\circ}{P}{}^{[2]}$ in Eq. (137), we obtain that

$$\frac{\partial}{\partial t}\overset{\circ}{P}{}^{[2]}(\mathbf{r}, t) = 2\tau_{n_2}^{-1}\eta(\mathbf{r}, t)\Lambda^{[2]}(\mathbf{r}, t) - \tau_{n_2}^{-1}\overset{\circ}{P}{}^{[2]}(\mathbf{r}, t), \quad (149)$$

in the absence of the external force, i.e., we set $\mathbf{u}_{ext} = 0$, and where $\eta(\mathbf{r}, t)$ is the shear viscosity coefficient of Eq. (146).

Finally, the evolution equation for the energy of Eq. (13) follows, after using Eq. (47) and the evolution equations for the velocity and the pressure, in the form

$$\frac{\partial}{\partial t}h(\mathbf{r}, t) = \frac{3}{2}\beta_B\tau_{n_2}^{-1}n(\mathbf{r}, t) - \frac{5}{2}\nabla \cdot [p(\mathbf{r}, t)\mathbf{v}(\mathbf{r}, t)]$$
$$- \frac{1}{2}\nabla \cdot \text{Tr}\{n(\mathbf{r}, t)[\mathbf{v}(\mathbf{r}, t)\mathbf{v}(\mathbf{r}, t)\mathbf{v}(\mathbf{r}, t)]\} - \tau_{n_2}^{-1}h(\mathbf{r}, t)$$
$$+ \frac{m}{2}\tau_{n_1}^{-1}\text{Tr}\{n(\mathbf{r}, t)[\mathbf{v}(\mathbf{r}, t)\mathbf{u}_{ext}(\mathbf{r}, t)]\}. \quad (150)$$

## V. FINAL COMMENTS

The formulation of a nonequilibrium thermodynamics and hydrodynamics, with an interplay between the macroscopic and microscopic levels of description, via a nonequilibrium statistical mechanics, currently represents a fundamental frontier in the physical sciences, involving not only the purely physical aspects, but also in what concerns technological–industrial applications.

In summary, in this paper it was derived a generalized hydrodynamics of fluids under driven flow and shear stress in the context of a non-equilibrium statistical thermodynamics based on a nonequilibrium statistical ensemble formalism. The nonequilibrium equations of state (which are coupled to the evolution of the basic variables that describe the hydrodynamic motion of the system) were derived. The information-theoretic entropy, its rate of change, and the quasi-free energy and quasi-chemical potential in nonequilibrium conditions were derived and discussed. Generalized diffusion-advection and Maxwell–Cattaneo advection equations were obtained in appropriate limiting situations. As an illustration, this nonlinear higher-order hydrodynamics was applied to the case of a dilute solution of Brownian particles in nonequilibrium conditions and flowing in a solvent acting as a thermal bath in the framework of such generalized hydrodynamics but truncated up to a second order.

Finally, we intend in a future study to take into account the two-particle distribution function [see Eq. (2) in Sec. II]. In this case, the generalized chemical potential depending on the pressure tensor could describe how a viscous pressure could modify the temperature of the phase transition.

## APPENDIX A: KINETIC FOUNDATIONS OF A GENERALIZED HYDRODYNAMICS

The construction of the nonlinear higher-order thermo-hydrodynamics here presented is based, as noticed, in a nonequilibrium statistical ensemble formalism (NESEF),[41,54,55] and for the sake of completeness, we here very briefly review its foundations. For such purpose, first it needs be noticed that for systems away from equilibrium, several important points need be carefully taken into account in each case under consideration:

1. The choice of the basic variables (a wholly different choice than in equilibrium when it suffices to take a set of those which are constants of motion), which is to be based on an analysis of what sort of macroscopic measurements and processes are actually possible, and, moreover, one is to focus attention not only on what can be observed but also on the character and expectative concerning the equations of evolution for these variables.[10,56] We also notice that even tough at the initial stage we would need to introduce all the observables of the system, and eventually variances, as time elapses more and more contracted descriptions can be used as it enters into play Bogoliubov's principle of correlation weakening and the accompanying hierarchy of relaxation times.[56]
2. The question of irreversibility (or Eddington's arrow of time) on what Peierls states that "In any theoretical treatment of transport problems, it is important to realize at what point the irreversibility has been incorporated. If it has not been incorporated, the treatment is wrong. A description of the situation that preserves the reversibility in time is bound to give the answer zero or infinity for any conductivity. If we do not see clearly where the irreversibility is introduced, we do not clearly understand what we are doing."[57]
3. Historicity needs be introduced, that is, the idea that it must be incorporated all the past dynamics of the system (or historicity effects), all along the time interval going from a starting description of the macro-state of the sample in the given experiment, say, at $t_0$, up the time $t$ when a measurement is performed. This is a quite important point in the case of dissipative systems as emphasized among others by Kirkwood and Mori.[58,59]

Concerning the question of the choice of the basic variables, differently to the case in equilibrium, immediately after the system of $N$ particles, in contact with external sources and reservoirs, has been driven out of equilibrium it would be necessary to describe its state in terms of all its observables and eventually, introducing direct and cross correlation fluctuations, but this is equivalent to have access to the so-called one-particle (or single particle), $\hat{n}_1$, and two-particle, $\hat{n}_2$, dynamical operators.

This is so because, we recall, all observable quantities and their variances can be expressed, at the microscopic mechanical level, in terms of these dynamical operators. For a description of mechanical states by means of these reduced density operators, we refer the reader to the already classical paper by Fano.[13]

For the sake of completeness, we notice that in classical mechanics the one-particle and two-particle operators, $\hat{n}_1$ and $\hat{n}_2$ are given, respectively, by

$$\hat{n}_1(\mathbf{r}, \mathbf{p}) = \sum_{j=1}^{N} \delta(\mathbf{r} - \mathbf{r}_j)\delta(\mathbf{p} - \mathbf{p}_j), \quad (A1)$$

$$\hat{n}_2(\mathbf{r}, \mathbf{p}, \mathbf{r}', \mathbf{p}') = \sum_{j \neq k=1}^{N} \delta(\mathbf{r} - \mathbf{r}_j)\delta(\mathbf{p} - \mathbf{p}_j)\delta(\mathbf{r}' - \mathbf{r}_k) \\ \times \delta(\mathbf{p}' - \mathbf{p}_k), \quad (A2)$$

where $\mathbf{r}_j$ and $\mathbf{p}_j$ are the coordinate and linear momentum of the $j$th particle in phase space and $\mathbf{r}_1, \mathbf{p}_1$, etc., the continuous values of position and momentum, which are sometimes called field variables (for simplicity we take the case of a single class of particles; otherwise, we must write dynamical operators for each kind of particles).

In quantum mechanics, the one- and two-particles density operators are

$$\hat{n}_1(\mathbf{r}, \sigma, \mathbf{r}', \sigma') = \psi_\sigma^\dagger(\mathbf{r})\psi_{\sigma'}(\mathbf{r}'), \quad (A3)$$

$$\hat{n}_2(\mathbf{r}_1, \sigma_1, \mathbf{r}_2, \sigma_2, \mathbf{r}'_2, \sigma'_2, \mathbf{r}'_1, \sigma'_1) = \psi_{\sigma_1}^\dagger(\mathbf{r}_1)\psi_{\sigma_2}^\dagger(\mathbf{r}_2)\psi_{\sigma'_2}(\mathbf{r}'_2)\psi_{\sigma'_1}(\mathbf{r}'_1), \quad (A4)$$

where $\sigma$ is the spin index and $\psi(\psi^\dagger)$ are single-particle field operators in second quantization (an excellent didactic description of them is available in the article by Robertson[60]).

Hence, on the one hand, the nonequilibrium statistical operator $\mathcal{R}_\varepsilon(t)$ is dependent on these quantities, and, on the other hand, the macro-variables for describing the nonequilibrium thermodynamic state of the system are the average value of the same quantities over the nonequilibrium ensemble, i.e.,

$$f_1(\mathbf{r}, \mathbf{p}; t) = \text{Tr}\{\hat{n}_1(\mathbf{r}, \mathbf{p})\mathcal{R}_\varepsilon(t)\}, \quad (A5)$$

$$f_2(\mathbf{r}_1, \mathbf{p}_1, \mathbf{r}_2, \mathbf{p}_2; t) = \text{Tr}\{\hat{n}_1(\mathbf{r}, \mathbf{p})\mathcal{R}_\varepsilon(t)\}, \quad (A6)$$

in classical mechanics; $f(\mathbf{r}, \mathbf{p}; t)$ has the role of a generalized Boltzmann distribution function.

On the question of irreversibility, Krylov[61] considered that there always exists a physical interaction between the measured system and the external world that is constantly "jolting" the system out of its exact microstate. Thus, the instability of trajectories and the unavoidable finite interactions with the outside would guarantee the working of a "crudely prepared" macroscopic description. In the absence of a proper way to introduce such effect, one needs to resort to the interventionist's approach, which is grounded on the basis of such ineluctable process of randomization leading to the asymmetric evolution of the macro-state.

The "intervention" consists in introducing into the Liouville equation of the statistical operator a particular source accounting for Krylov's "jolting" effect, in the form (written for the logarithm of the statistical operator),

$$\frac{\partial}{\partial t}\ln \mathcal{R}_\varepsilon(t) + \frac{1}{i\hbar}\left[\ln \mathcal{R}_\varepsilon(t), \hat{H}\right] = -\varepsilon\left[\ln \mathcal{R}_\varepsilon(t) - \ln \bar{\mathcal{R}}(t, 0)\right], \quad (A7)$$

where $\varepsilon$ (kind of reciprocal of a relaxation time) is taken to go to $+0$ after the calculation of average values is performed. Such mathematically inhomogeneous term, in the otherwise normal Liouville equation, implies in a continuous tendency of relaxation of the statistical operator toward a referential one, $\bar{\mathcal{R}}$, which, as discussed below, represents an instantaneous quasi-equilibrium condition.

We can see that Eq. (A7) consists of a regular Liouville equation but with an infinitesimal source, which introduces Bogoliubov's symmetry breaking of time reversal and is responsible for disregarding the advanced solutions.[27,62] This is described by a Poisson distribution and the result at time $t$ is obtained by averaging over all $t'$ in the interval $(t_0, t)$, once the solution of Eq. (A7) is

$$\mathcal{R}_\varepsilon(t) = \exp\left\{-\hat{S}(t, 0) + \int_{t_0}^{t} dt' \omega_\varepsilon(t', t) \frac{d}{dt'}\hat{S}(t', t' - t)\right\}, \quad (A8)$$

where
$$\hat{S}(t,0) = -\ln \bar{\mathcal{R}}(t,0), \quad (A9)$$

$$\hat{S}(t', t'-t) = \exp\left\{-\frac{1}{i\hbar}(t'-t)\hat{H}\right\}\hat{S}(t',0)$$
$$\times \exp\left\{\frac{1}{i\hbar}(t'-t)\hat{H}\right\}, \quad (A10)$$

and $\omega_\varepsilon$ is the Poissonian weight function[29,63]
$$\omega_\varepsilon(t',t) = \varepsilon \exp\{\varepsilon(t'-t)\}, \quad (A11)$$

and the initial condition at time $t_0$, when the formalism begins to be applied, is
$$\mathcal{R}_\varepsilon(t_0) = \bar{\mathcal{R}}(t_0, 0). \quad (A12)$$

This time $t_0$, of initiation of the statistical description, is usually taken in the remote past ($t_0 \to -\infty$) introducing an adiabatic switch on of the relaxation processes, and in Eq. (A8) the interaction in time in the interval $(t_0, t)$ is weighted by the kernel $\omega_\varepsilon(t',t)$ of Eq. (A11). The presence of this kernel introduces a kind of "evanescent history" as the system macro-state evolve toward the future from the boundary condition of Eq. (A12) at time ($t_0 \to -\infty$) [a result of the presence of the exponential in Eq. (A8), which accounts for the dissipative evolution of the state of the system, a fact evidenced in the resulting kinetic theory[27] which clearly indicates that a fading memory process has been introduced]. Moreover, in most cases we can consider the system as composed of the system of interest (on which we are performing an experiment) in contact with ideal reservoirs. Thus, we can write
$$\mathcal{R}_\varepsilon(t) = \varrho_\varepsilon(t) \times \varrho_R, \quad (A13)$$

where $\varrho_\varepsilon(t)$ is the statistical operator of the nonequilibrium system and $\varrho_R$ the stationary on of the ideal reservoirs, with $\varrho_\varepsilon(t)$ given then by
$$\varrho_\varepsilon(t) = \exp\left\{-\hat{S}(t,0) + \int_{-\infty}^{t} dt' e^{\varepsilon(t'-t)} \frac{d}{dt'}\hat{S}(t',t'-t)\right\}, \quad (A14)$$

having the initial value $\bar{\varrho}(t_0, 0)$ ($t_0 \to -\infty$), and where
$$\hat{S}(t,0) = -\ln \bar{\varrho}(t,0). \quad (A15)$$

Finally, it needs to be provided the auxiliary statistical operator $\bar{\varrho}(t,0)$. It defines an instantaneous distribution at time $t$, which describes a "frozen" equilibrium by providing at such given time the macroscopic state of the system, and for that reason is sometimes dubbed as the quasi-equilibrium statistical operator. On the basis of this (or, alternatively, via the extremum principle procedure[27]), and considering the description of the nonequilibrium state of the system in terms of the single- and two-particle density operators, the reference or instantaneous quasi-equilibrium statistical operator is taken as a canonical-like one given by
$$\bar{\varrho}(t,0) = \exp\left\{-\phi(t) - \int d^3r \int d^3p F_1(\mathbf{r},\mathbf{p};t)\hat{n}_1(\mathbf{r},\mathbf{p})\right.$$
$$\left. - \int d^3r \int d^3p \int d^3r' \int d^3p' F_2(\mathbf{r},\mathbf{p},\mathbf{r}',\mathbf{p}';t) \times \hat{n}_2(\mathbf{r},\mathbf{p},\mathbf{r}',\mathbf{p}')\right\},$$
$$(A16)$$

in the classical case, with $\phi(t)$ ensuring the normalization of $\bar{\varrho}(t,0)$, and playing the role of a kind of a logarithm of a partition function, say, $\phi(t) = \ln \bar{Z}(t)$. Moreover, in this Eq. (A16), $F_1$ and $F_2$, are the nonequilibrium thermodynamic variables associated with each kind of basic dynamical variables $\hat{n}_1$ and $\hat{n}_2$, respectively (Lagrange multipliers in the extremum principle approach).

An alternative equivalent and complete descriptive, highly convenient for, as shown here, deriving a kinetic theory of hydrodynamics (a nonlinear higher-order one), consists in the construction of a generalized nonequilibrium grand-canonical distribution. Considering for simplicity the case of retaining only $\hat{n}$, as basic dynamical variable, this follows by redefining the nonequilibrium thermodynamic variable, in the form
$$F_1(\mathbf{r},\mathbf{p};t) = F_n(\mathbf{r},t) + \frac{\mathbf{p}}{m} \cdot \mathbf{F}_n(\mathbf{r},t) + \frac{p^2}{2m}F_h(\mathbf{r},t) + \frac{p^2}{2m}\frac{\mathbf{p}}{m} \cdot \mathbf{F}_h(\mathbf{r},t)$$
$$+ \sum_{i \geq 2}\left[F_h^{[\ell]}(\mathbf{r},t) \otimes \frac{p^2}{2m}u_n^{[\ell]}(\mathbf{p}) + F_n^{[\ell]}(\mathbf{r},t) \otimes u_n^{[\ell]}(\mathbf{p})\right],$$
$$(A17)$$

where
$$u_n^{[\ell]}(\mathbf{p}) = \left[\frac{\mathbf{p}}{m}\ldots(\ell - \text{times})\ldots\frac{\mathbf{p}}{m}\right] \quad (A18)$$

is a $\ell$-rank tensor consisting of the tensorial product of $\ell$-times the velocity $\mathbf{p}/m$, and $\otimes$ stands for fully contracted product of tensors, which introduced in the statistical operator of Eq. (A17) has taken the form
$$\bar{\varsigma}(t,0) = \exp\left\{-\phi(t) - \int d^3r \left[F_h(\mathbf{r},t)\hat{h}(\mathbf{r}) + F_n(\mathbf{r},t)\hat{n}(\mathbf{r})\right.\right.$$
$$+ \mathbf{F}_h(\mathbf{r},t) \cdot \hat{\mathbf{I}}_h(\mathbf{r}) + \mathbf{F}_n(\mathbf{r},t) \cdot \hat{\mathbf{I}}_n(\mathbf{r})\right]$$
$$\left.+ \sum_{\ell \geq 2}\left[F_h^{[\ell]}(\mathbf{r},t) \otimes \hat{I}_h^{[\ell]}(\mathbf{r}) + F_n^{[\ell]}(\mathbf{r},t) \otimes \hat{I}_n^{[\ell]}(\mathbf{r},t)\right]\right\},$$
$$(A19)$$

where
$$\hat{h}(\mathbf{r}) = \int d^3p \frac{p^2}{2m}\hat{n}_1(\mathbf{r},\mathbf{p}), \quad (A20)$$

$$\hat{n}(\mathbf{r}) = \int d^3p \, \hat{n}_1(\mathbf{r},\mathbf{p}), \quad (A21)$$

$$\hat{\mathbf{I}}_h(\mathbf{r}) = \int d^3p \frac{\mathbf{p}}{2m}\frac{p^2}{2m}\hat{n}_1(\mathbf{r},\mathbf{p}), \quad (A22)$$

$$\hat{\mathbf{I}}_n(\mathbf{r}) = \int d^3p \frac{\mathbf{p}}{m}\hat{n}_1(\mathbf{r},\mathbf{p}), \quad (A23)$$

$$\hat{I}_h^{[\ell]}(\mathbf{r}) = \int d^3p \, u^{[\ell]}(\mathbf{p})\frac{p^2}{2m}\hat{n}_1(\mathbf{r},\mathbf{p}), \quad (A24)$$

$$\hat{I}_n^{[\ell]}(\mathbf{r}) = \int d^3p \, u^{[\ell]}(\mathbf{p})\hat{n}_1(\mathbf{r},\mathbf{p}), \quad (A25)$$

which are the densities of energy, $\hat{h}$, and particles, $\hat{n}$ and their fluxes of all order (the vectorial ones and the tensorial ones with $\ell \geq 2$). They have as conjugated nonequilibrium thermodynamic variables the set

$$\left\{F_h(\mathbf{r},t), F_n(\mathbf{r},t), \mathbf{F}_h(\mathbf{r},t), \mathbf{F}_n(\mathbf{r},t), \{F_h^{[2]}(\mathbf{r},t)\}, \{F_n^{[2]}(\mathbf{r},t)\}\right\},$$
(A26)

and it can be noticed that, alternatively, this set of variables completely describes the nonequilibrium thermodynamic state of the system. They are related to the basic set of macro-variables by the relations (which are the equivalent of equations of state in arbitrary nonequilibrium conditions[2]),

$$\mathbf{I}_h^{[\ell]}(\mathbf{r},t) = \mathrm{Tr}\{\hat{\mathbf{I}}_h(\mathbf{r})\bar{\varrho}(t,0)\} = \frac{\delta\phi(t)}{\delta F_h^{[\ell]}(\mathbf{r},t)} = \frac{\delta \ln \bar{Z}(t)}{\delta F_h^{[\ell]}(\mathbf{r},t)}, \quad \text{(A27)}$$

$$\mathbf{I}_n^{[\ell]}(\mathbf{r},t) = \mathrm{Tr}\{\hat{\mathbf{I}}_n(\mathbf{r})\bar{\varrho}(t,0)\} = \frac{\delta\phi(t)}{\delta F_n^{[\ell]}(\mathbf{r},t)} = \frac{\delta \ln \bar{Z}(t)}{\delta F_n^{[\ell]}(\mathbf{r},t)}, \quad \text{(A28)}$$

where $\ell = 0$ for the densities, $\ell = 1$ for the vector (first order) fluxes, and $\ell \geq 2$ for the higher-order tensor fluxes, we have used that $\varrho_\varepsilon$ and $\bar{\varrho}$, at each time $t$, define the same average values for the basic variables only,[27] and $\delta$ stands for functional differential.[64] A complete description of the NESEF foundations for irreversible thermodynamical is available in Ref. 2. Finally, we noticed that the NESEF-based nonlinear kinetic theory of relaxation processes basically consists in taking the average over the nonequilibrium ensemble of Heisenberg (or Hamilton at the classical level) equations of motion of the dynamical operators for the observable, say, $\hat{A}$, under consideration, i.e.,

$$\frac{\partial}{\partial t}\mathrm{Tr}\{\hat{A}(\mathbf{r})\varrho_\varepsilon(t)\} = \frac{1}{i\hbar}\mathrm{Tr}\{[\hat{A}(\mathbf{r}),\hat{H}]\varrho_\varepsilon(t)\}, \quad \text{(A29)}$$

which is a manifestation of Ehrenfest theorem. The practical handling of this NESEF-kinetic theory is described in Ref. 41 with applications in Refs. 65–70.

## APPENDIX B: DERIVATION OF THE NONEQUILIBRIUM EQUATIONS OF STATE

Taking into account that

$$\psi(t) = \int \frac{d^3r\, d^3p}{(2\pi\hbar)^3} e^{-\Psi(\mathbf{r},t)}, \quad \text{(B1)}$$

with

$$\Psi(\mathbf{r},t) = \beta^*(\mathbf{r},t)\frac{p^2}{2m} - \mu^*(\mathbf{r},t) - \mathbf{v}_n(\mathbf{r},t)\cdot\mathbf{p}$$
$$- \frac{\beta^*(\mathbf{r},t)}{m}v_n^{[2]}(\mathbf{r},t) \times [\mathbf{pp}], \quad \text{(B2)}$$

introducing the shift $\mathbf{p} \to \mathbf{p} - \boldsymbol{\kappa}(\mathbf{r},t)$ with

$$M^{[2]}(\mathbf{r},t)\boldsymbol{\kappa}(\mathbf{r},t) = m\beta^*(\mathbf{r},t)v_n(\mathbf{r},t), \quad \text{(B3)}$$

where

$$M^{[2]}(\mathbf{r},t) = \beta^*(\mathbf{r},t)\mathbf{1}^{[2]} + \frac{2}{m}F_n^{[2]}(\mathbf{r},t) = 0 \quad \text{(B4)}$$

[with $F_n^{[2]}(\mathbf{r},t)$ of Eq. (92)], what is done in order to eliminate the linear term in $\mathbf{p}$ in the integrand in Eq. (B1), we arrive at that

$$\psi(t) = \left(\frac{m}{2\pi\hbar^2}\right)^{2/3} \int d^3r\, \frac{e^{-\xi(\mathbf{r},t)}}{\sqrt{M(\mathbf{r},t)}}, \quad \text{(B5)}$$

where $M$ is the determinant of $M^{[2]}$ and

$$\xi(\mathbf{r},t) = \beta^*(\mathbf{r},t)\left(\frac{\kappa^2(\mathbf{r},t)}{2m} - \mu^*(\mathbf{r},t) - \mathbf{v}_n(\mathbf{r},t)\cdot\boldsymbol{\kappa}(\mathbf{r},t)\right)$$
$$- \frac{1}{m^2}F_n^{[2]}(\mathbf{r},t) \otimes [\boldsymbol{\kappa}(\mathbf{r},t)\boldsymbol{\kappa}(\mathbf{r},t)]. \quad \text{(B6)}$$

We are now in conditions to obtain the nonequilibrium equations of state, that is, the relation between the basic thermo-hydrodynamical variables and the nonequilibrium thermodynamic variables. They follow calculating Eqs. (27)–(29) or alternatively can be obtained from the knowledge of the nonequilibrium partition function,

$$\ln \bar{Z}(t) = e^{\psi(t)},$$

via the functional differentials[64]

$$n(\mathbf{r},t) = -\frac{\delta\psi(t)}{\delta A_n(\mathbf{r},t)}, \quad \text{(B7)}$$

$$h(\mathbf{r},t) = -\frac{\delta\psi(t)}{\delta F_h(\mathbf{r},t)}, \quad \text{(B8)}$$

$$\mathbf{I}(\mathbf{r},t) = -\frac{\delta\psi(t)}{\delta \mathbf{F}_n(\mathbf{r},t)}, \quad \text{(B9)}$$

$$I^{[2]}(\mathbf{r},t) = -\frac{\delta\psi(t)}{\delta F_n^{[2]}(\mathbf{r},t)}. \quad \text{(B10)}$$

The corresponding expressions are

$$n(\mathbf{r},t) = \left(\frac{m}{2\pi\hbar^2}\right)^{2/3} \frac{e^{\beta^*(\mathbf{r},t)}}{\sqrt{M(\mathbf{r},t)}}, \quad \text{(B11)}$$

with

$$\beta^*(\mathbf{r},t)\tilde{\mu}(\mathbf{r},t) = \beta^*(\mathbf{r},t)\mu^*(\mathbf{r},t) - \frac{1}{2}mv^2(\mathbf{r},t)$$
$$- \frac{1}{2}mM^{[2]}(\mathbf{r},t) \otimes [\mathbf{v}(\mathbf{r},t)\mathbf{v}(\mathbf{r},t)], \quad \text{(B12)}$$

$$h(\mathbf{r},t) = \frac{1}{2}n(\mathbf{r},t)\mathrm{Tr}\{\mathcal{M}^{[2]}(\mathbf{r},t)\} + n(\mathbf{r},t)\frac{1}{2m}\kappa^2(\mathbf{r},t), \quad \text{(B13)}$$

and $\mathcal{M}^{[2]}$ being the inverse of $M^{[2]}$,

$$\mathbf{I}(\mathbf{r},t) = \frac{1}{m}n(\mathbf{r},t)\boldsymbol{\kappa}(\mathbf{r},t) \equiv n(\mathbf{r},t)\mathbf{v}(\mathbf{r},t), \quad \text{(B14)}$$

after defining the velocity

$$\mathbf{v}(\mathbf{r},t) = \frac{\boldsymbol{\kappa}(\mathbf{r},t)}{m}, \quad \text{(B15)}$$

which corresponds to the barycentric velocity of standard hydrodynamics, and finally

$$I_n^{[2]} = n(\mathbf{r},t)\left(\frac{1}{m}\mathcal{M}^{[2]}(\mathbf{r},t) + \frac{1}{m^2}[\boldsymbol{\kappa}(\mathbf{r},t)\boldsymbol{\kappa}(\mathbf{r},t)]\right),$$
$$I_n^{[2]} = n(\mathbf{r},t)\left(\frac{1}{m}\mathcal{M}^{[2]}(\mathbf{r},t) + [\mathbf{v}(\mathbf{r},t)\mathbf{v}(\mathbf{r},t)]\right). \quad \text{(B16)}$$

It can also be noticed that, according to Eqs. (B3), (B15), and (49), the barycentric velocity $\mathbf{v}(\mathbf{r},t)$ and the drift velocity $\mathbf{v}_n(\mathbf{r},t)$ are related by the expression

$$n(\mathbf{r},t)\mathbf{v}(\mathbf{r},t) = n(\mathbf{r},t)\frac{\boldsymbol{\kappa}(\mathbf{r},t)}{m},$$
$$n(\mathbf{r},t)\mathbf{v}(\mathbf{r},t) = -n(\mathbf{r},t)\beta^*(\mathbf{r},t)\mathcal{M}^{[2]}(\mathbf{r},t)\mathbf{v}_n(\mathbf{r},t), \quad \text{(B17)}$$
$$n(\mathbf{r},t)\mathbf{v}(\mathbf{r},t) = -P^{[2]}(\mathbf{r},t)\beta^*(\mathbf{r},t)\mathbf{v}_n(\mathbf{r},t).$$

## APPENDIX C: THE COLLISION INTEGRALS

### A. The collision integral associated with the flux

Let us consider the collision integral present in Eq. (90), i.e., the evolution equation for the flux of particles. It is given, in the Markovian approximation to the NESEF-based kinetic theory, by

$$\mathbf{J}_n(\mathbf{r},t) = \mathbf{J}_n^{(1)}(\mathbf{r},t) + \mathbf{J}_n^{(2)}(\mathbf{r},t) + \mathbf{J}_n^{(3)}(\mathbf{r},t). \quad \text{(C1)}$$

After performing the calculation, it follows that

$$\mathbf{J}_n^{(1)}(\mathbf{r},t) = -A^{[2]}(\mathbf{r},t)\mathbf{I}(\mathbf{r},t), \quad \text{(C2)}$$

with

$$A^{[2]}(\mathbf{r},t) = \frac{M_R}{M_x}\sqrt{\frac{\pi(M\beta_R)^3}{2}} \times \sum_{\mathbf{Q}} \frac{|\Psi(\mathbf{Q})|^2}{Q}[\mathbf{QQ}]$$
$$\times \left[ \mathbf{1}^{[2]} - \frac{M\beta_R}{mQ^2}[\mathbf{QQ}] \times \left(M^{[2]}(\mathbf{r},t) + \frac{M\beta_R}{mQ^2}[\mathbf{QQ}]\right)^{-1}\right] f(\mathbf{Q},t), \quad \text{(C3)}$$

being a kind of reciprocal of a Maxwell's characteristic time tensor, where $|\Psi(\mathbf{Q})|^2$ is the modulus square of the Fourier transform of the interaction potential, $M$ the mass of particles in the reservoir, $\beta_R = 1/k_B T_R$, and

$$f(\mathbf{Q},t) = \sqrt{M(\mathbf{r},t)} \times \left[\det\left(M^{[2]}(\mathbf{r},t) + \frac{M\beta_R}{m_x Q^2}[\mathbf{QQ}]\right)\right]^{-1/2}$$
$$\times \exp\left\{\frac{1}{2m}\left(M^{[2]}(\mathbf{r},t) + \frac{M\beta_R}{m_x Q^2}[\mathbf{QQ}]\right) \otimes [\mathbf{q}(\mathbf{r},t)\mathbf{q}(\mathbf{r},t)]\right.$$
$$\left. - \frac{M\beta_R}{2Q^2}[\mathbf{QQ}] \otimes [\mathbf{v}(\mathbf{r},t)\mathbf{v}(\mathbf{r},t)]\right\}, \quad \text{(C4)}$$

recalling that $n(\mathbf{r},t)\mathbf{v}(\mathbf{r},t) = \mathbf{I}(\mathbf{r},t)$ and where $\mathbf{q}$ is defined by

$$\left(M^{[2]}(\mathbf{r},t) + \frac{M\beta_R}{m_x Q^2}[\mathbf{QQ}]\right)\mathbf{q}(\mathbf{r},t) = -\frac{M\beta_R}{Q^2}[\mathbf{QQ}]\mathbf{v}(\mathbf{r},t). \quad \text{(C5)}$$

Finally,

$$n_R = N_R/V$$

and

$$m_x^{-1} = m^{-1} + M^{-1}.$$

Hence, in Eq. (C1) we do have the contribution to the scattering operator of the form of Maxwell's one, but two other contributions are present: we do no enter the cumbersome expressions form then, suffice to say that $J_n^{(2)}$ contains the divergence of the pressure tensor (the divergence of the second order flux) and the gradient of the concentration, as it should according to mesoscopic irreversible thermodynamics:[54] These terms, as shown in Ref. 33, lead to contributions of Burnett and super-Burnett type. $J_n^{(3)}$ is a nonlocal contribution, containing space statistical higher-order hydrodynamics presented in Refs. 71 and 72.

For the case of Brownian particles, when $m/M \ll 1$, Eq. (C1) gives over the simple form

$$\mathbf{J}_n(\mathbf{r},t) = -\varsigma_{n_1}^{-1}\mathbf{I}_n(\mathbf{r},t), \quad \text{(C6)}$$

where

$$\varsigma_{n_1}^{-1} = \frac{1}{3}\sqrt{\frac{\pi(\beta_R)^3}{2m}} n_R \mathcal{F}(0), \quad \text{(C7)}$$

with

$$\mathcal{F}(0) = \frac{1}{V}\sum_{\mathbf{q}} q|v(\mathbf{q})|^2. \quad \text{(C8)}$$

### B. The collision integral associated with the second flux

Performing the calculations along similar line as done in Appendix C A, we obtain in the case of the Brownian particles that

$$J_n^{[2]}(\mathbf{r},t) = -\varsigma_{n_2}^{-1} I_n^{[2]}(\mathbf{r},t), \quad \text{(C9)}$$

where

$$\varsigma_{n_2}^{-1} = 2\varsigma_{n_1}^{-1}. \quad \text{(C10)}$$

## DATA AVAILABILITY

The data that support the findings of this study are available within the article.